\documentclass[11pt,a4paper]{article}
\pdfoutput=1
\usepackage{jheppub}
\usepackage{amsthm,amsbsy,amsfonts,mathrsfs,enumerate,float,wrapfig,amsmath,amssymb,bm}
\usepackage[colorinlistoftodos,prependcaption,textsize=scriptsize]{todonotes}

\newcommand{\be}{\begin{equation}}
\newcommand{\ee}{\end{equation}}
\newcommand{\bea}{\begin{eqnarray}}
\newcommand{\eea}{\end{eqnarray}}
\newcommand{\Emptyset}{\text{\o}}
\allowdisplaybreaks

\title{Surface defects on E-string from 5-brane webs}

\author[a]{Sung-Soo Kim,}
\author[b,c]{Yuji Sugimoto,}
\author[d]{and Futoshi Yagi}

\affiliation[a]{School of Physics, University of Electronic Science and Technology of China, \\
No.2006 Xiyuan Ave, West Hi-Tech Zone, 
Chengdu, Sichuan 611731, China}
\affiliation[b]{Interdisciplinary Center for Theoretical  Study, University of Science and Technology of China,\\
 96 Jinzai Road, Hefei, Anhui 230026, China}
\affiliation[c]{Peng Huanwu Center for Fundamental Theory,\\
 96 Jinzai Road, Hefei, Anhui 230026, China}
\affiliation[d]{School of Mathematics, Southwest Jiaotong University,\\ 
West zone, High-tech district, Chengdu, Sichuan 611756, China}

\emailAdd{sungsoo.kim@uestc.edu.cn}
\emailAdd{sugimoto@ustc.edu.cn}
\emailAdd{futoshi\_yagi@swjtu.edu.cn}

\abstract{
We study 6d E-string theory with defects on a circle. Our basic strategy is to apply the geometric transition to the supersymmetric gauge theories. First, we calculate the partition functions of the 5d SU(3)$_0$ gauge theory with 10 flavors, which is UV-dual to the 5d Sp(2) gauge theory with 10 flavors, based on two different 5-brane web diagrams, and check that two partition functions agree with each other. Then, by utilizing the geometric transition, we find the surface defect partition function for E-string on $\mathbb{R}^4\times T^2$. We also discuss that our result is consistent with the elliptic genus. Based on the result, we show how the global symmetry is broken by the defects, and discuss that the breaking pattern depends on where/how we insert the defects.
}

\begin{document}
\preprint{
USTC-ICTS/PCFT-20-23
}

\maketitle


\section{Introduction}\label{sec:intro}

Study of surface defects \cite{Gukov:2006jk,Gukov:2008sn} provides us with a tool for uncovering novel structure of non-perturbative aspects as well as enriching interplay between theories of codimension 2 \cite{Gomis:2007fi, Alday:2009fs, Alday:2010vg, Gaiotto:2012xa, Gaiotto:2013sma, Gaiotto:2014ina, Nazzal:2018brc}.
A way to see how the defects affect physical systems is to calculate the partition functions.
It is known that partition functions for some class of 5d $\mathcal{N}=1$ supersymmetric gauge theories compactified on a circle are equivalent to topological string amplitudes on corresponding non-compact toric Calabi--Yau manifolds under suitable parameter correspondence, which is known as geometric engineering \cite{Katz:1996fh, Katz:1997eq, Dijkgraaf:2002fc, Hollowood:2003cv}. Based on the equivalence between toric Calabi-Yau manifold and 5-brane web \cite{Leung:1997tw}, we can translate the geometry into brane set up and vice versa.

In the presence of the defects, the partition function can be computed by implementing 5d system with the defects on a Type IIB 5-brane web, where the defects are realized as (perpendicular) D3-branes inserted on the $(p,q)$-web plane for supersymmetric gauge theories. 
Such defect insertion is captured as a particular choice of K\"ahler parameters in topological string amplitudes, which corresponds to open topological string amplitudes \cite{Dimofte:2010tz, Awata:2010bz}, which is known as geometric transition~\cite{Gopakumar:1998ki, Ooguri:1999bv}.

The procedure of obtaining the defect partition functions can be viewed as a generalization of Higgsing procedure from 5-brane webs. Depending how to choose or tune the K\"ahler parameters, one sees usual Higgsing or a system with defect.
More precisely, to obtain the defect partition function, we need to tune K\"ahler parameters in a way that not only it reduces the rank of gauge group but also yields the open topological string partition function up to Coulomb branch independent overall factors such as MacMahon function or the extra factors. Geometric transition when the number of inserted D3 defects is zero, reduces to usual Higgsing. In this regard, it is a generalization of Higgsing procedure, and we refer to it as defect Higgsings. Using the defect Higgsing, the 5d defect partition functions have been computed in a straightforward manner.

Many 6d theories on a circle are also realized on 5-brane webs, one can hence also apply the defect Higgsing to such KK theories. One example is M-string theory with surface defects inserted \cite{Mori:2016qof}, where M-string is realized as a periodic $(p,q)$ 5-brane webs where the NS5-branes are identified. This gives rise to a web on a cylinder where the circular direction corresponds to 
the compactification direction of 6d theories. 
\cite{Haghighat:2013gba,Sugimoto:2015nha}.

As another example of defect Higgsing, in this paper, we study 6d E-string theory with surface defects. E-string theory on a circle of radius $R$ is realized as SU(2) gauge theory with 8 hypermultiplets in the fundamental representation (flavors). There are two different 5-brane webs for the SU(2) gauge theory. One is of a spirally periodic shape whose period is identified as $R^{-1}$, which is called Tao diagram \cite{Kim:2015jba}. The other is of two O5-planes with their distance $R^{-1}$. Since an orientifold plane is used to realize Sp or SO gauge group in general, we refer to the former 5-brane web as web diagram for SU(2) gauge theory, while the latter as web diagram for Sp(1) gauge theory in this paper even though SU(2) and Sp(1) are identical. Partition function can be computed based on 5-brane webs via topological vertex \cite{Aganagic:2003db}. In particular, topological vertex formalism in the presence of O5-planes is also developed in \cite{Kim:2017jqn} and both topological string partition functions agree with the elliptic genus of 6d E-string theory.

Our strategy of computing the E-string partition function with defects is to first consider 5d SU(3)$_0$ gauge theory with 10 flavors and then apply the defect Higgsing. The 5d SU(3)$_0$ gauge theory with 10 flavors is UV-dual to the 5d Sp(2) theory with 10 flavors, in the sense that their UV completion is the same \cite{Gaiotto:2015una, Hayashi:2015fsa}. The corresponding 6d theory description is given by 6d Sp(1) gauge theory with 10 flavors and a tensor multiplet. Together with the brane configurations, these dual descriptions allow one to compute the partition function in several different ways. For instance, the ADHM-like method ~\cite{Yun:2016yzw}, the elliptic genus~\cite{Kim:2014dza}, and the topological vertex method~\cite{Hayashi:2016abm}. We compute the defect partition function by using the topological vertex method and compare it with the elliptic genus for E-string with defects which is obtained from 6d Sp(1) gauge theory with 10 flavors.

Analogous to the case with the 5d SU(2) gauge theory, the 5d SU(3)$_0$ gauge theory with 10 flavors has two different 5-brane configurations: One is without O5-plane and the other is with two O5-planes. The 5-brane configuration without O5-plane is Tao diagram \cite{Kim:2015jba}. This Tao diagram can be obtained from the 5-brane web diagram with two O7$^-$-planes, which is T-dual of the type IIA brane setup with an O8$^-$-plane \cite{Brunner:1997gf, Hanany:1997gh} for 6d Sp(1) gauge theory with 10 flavors and a tensor multiplet. By resolving both O7$^-$-planes into two different 7-branes~\cite{Sen:1996vd}, respectively, we obtain the diagram for the 5d SU(3)$_0$ gauge theory, which can be deformed to be a Tao diagram. On the other hand, if we resolve only one O7$^-$-plane, we obtain the diagram for the 5d Sp(2) gauge theory, which explains the UV-duality between the SU(3) gauge theory and the Sp(2) gauge theory \cite{Gaiotto:2015una, Hayashi:2015fsa, Hayashi:2016abm}. The 5-brane configuration with two O5-planes is again the T-dual of the type IIA brane setup but with O6-plane instead of O8-plane. As abuse of notation, we refer to the former web diagram without O5-planes as the web diagram for SU(3) gauge theory, while the latter diagram with two O5-planes as the web diagram for Sp(2) gauge theory in this paper, based on the knowledge that orientifold planes are used to realize Sp or SO gauge group in general.
\footnote{
Rigorously speaking, it would be more proper to understand that we obtain 5d SU(3) gauge theory or 5d Sp(2) gauge theory depending on the parameter region of the Wilson lines introduced to the 6d superconformal theory realized at the UV fixed point as well as the compactification radius. It indicates that both gauge theories can be realized in either of the two 5-brane web configurations. However, we use this notation just for simplicity.
}

The organization of the paper is as follows. In section \ref{sec:equiv}, we review the partition function for 5d SU(3)$_0$ gauge theory with 10 flavors from two different 5-branes setups: one with a 5-brane web without O5-planes and the other with two O5-planes. In section \ref{sec:estring}, using 5d 5-brane configurations for the SU(3)$_0$ gauge theory with 10 flavors, we perform defect Higgsing to yields 5d SU(2) gauge theory with 8 flavors. For comparison, we implement the defect Higgsing to the elliptic genus partition function of E-string theory to the agreement. In section \ref{sec:globsymm}, we discuss some issues of global symmetry in the presence of defects. We then conclude and discuss unbroken global symmetry and possible generalizations. In Appendix, we discuss our conventions, decoupling limit to get 5d theories from KK theories, and defects on pure SU(2) theories with different discrete theta angles.
\bigskip
\section{5d SU(3) gauge theories, 5-brane webs, and UV duality}\label{sec:equiv}

In this section, we consider the 5d $\mathcal{N}=1$ SU(3)$_\kappa$ gauge theories from the perspective of 5-brane webs in Type IIB string theory. In particular, we discuss how to obtain the BPS partition function of the SU(3)$_0$ gauge theory with 10 flavors of the Chern-Simons level $\kappa=0$. 
The computation is performed based on two different 5-brane web diagrams.
One is a 5-brane web diagram without O5-planes, which is Tao diagram introduced in \cite{Kim:2015jba}.
Though it is spirally periodic, one can apply the topological vertex method to compute the unrefined Nekrasov partition function. 
The other is a 5-brane web with two O5-planes. 
It is also possible to implement the topological vertex method to the 5-brane configurations with O5-plane(s) with special deformation of web diagrams and careful Young diagram assignments near O5-planes \cite{Kim:2017jqn}.

The partition function for 5d SU(2) with 8 flavors has been already computed explicitly  \cite{Hwang:2014uwa,Kim:2015jba,Kim:2017jqn}. The resulting partition functions based on a Tao diagram \cite{Kim:2015jba} and a 5-brane with two O5-planes \cite{Kim:2017jqn} look quite different, however, they agree up to unphysical factors, called {\it extra factor}, which do not depend on the Coulomb branch parameters.
As for the 5d $\mathcal{N}=1$ SU(3)$_0$ gauge theory with 10 flavors,
the topological string partition function based on a Tao diagram has been already computed \cite{Hayashi:2016abm}, but it has not been computed explicitly based on a 5-brane web with two O5-planes yet. 
In section \ref{sec:PartTao}, we first review the computation with the topological vertex based on the Tao diagram. 
Then, in section \ref{sec;equivO5}, we compute the partition function based on the 5-brane web with two O5-planes.
We refer to the former partition function as the partition function for 5d $\mathcal{N}=1$ SU(3)$_0$ gauge theory with 10 flavors, while the latter one as the partition function for 5d $\mathcal{N}=1$ Sp(2) gauge theory with 10 flavors.
We will see agreement between them by expressing the latter partition function in terms of the parameters of the SU(3) gauge theory.
The detailed computations and some notations are summarized in Appendix \ref{sec:appendix}.

\subsection{Partition function from Tao diagram}\label{sec:PartTao}
\begin{figure}[htb]
\centering
\includegraphics[width=15cm]{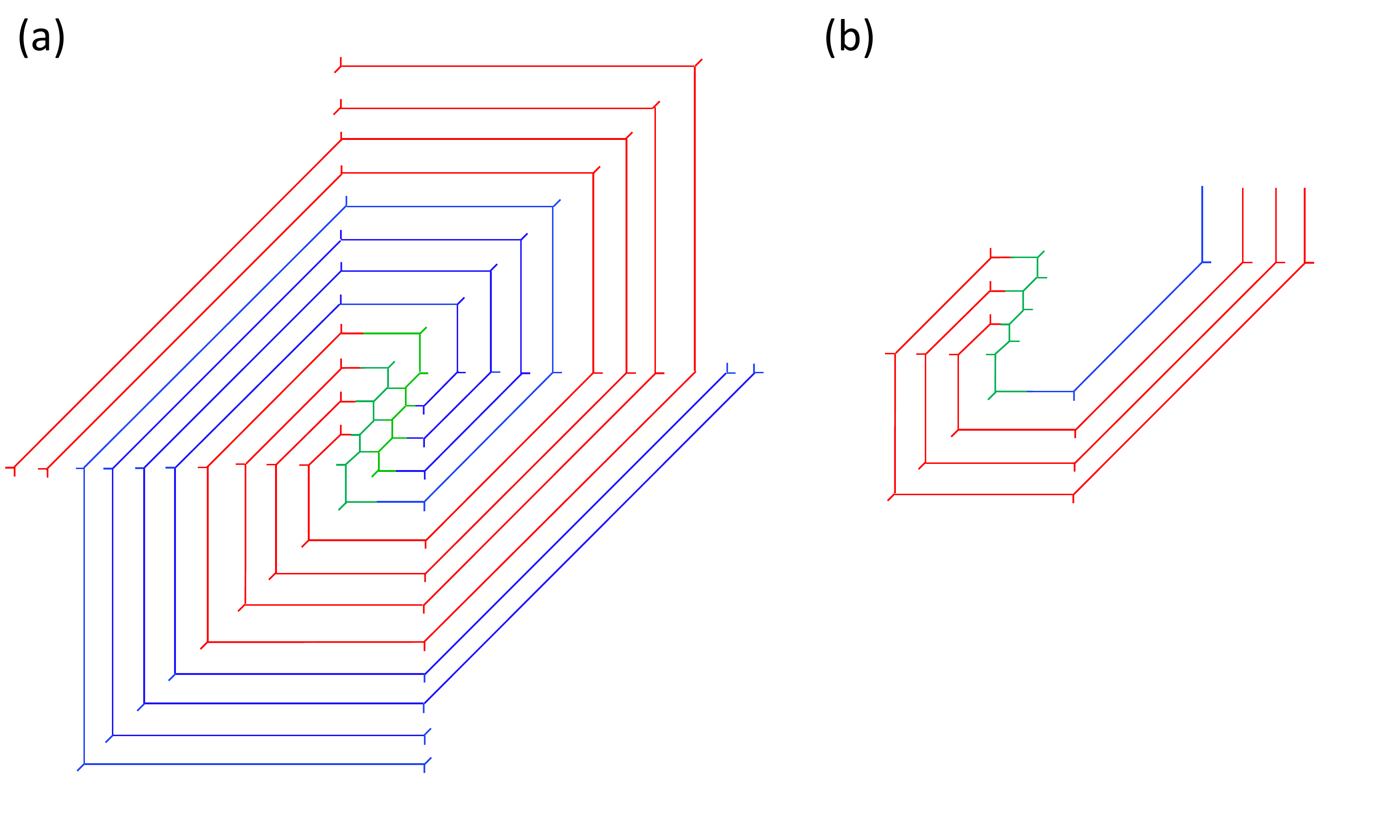}
\caption{(a) The Tao diagram corresponding to the 5d $\mathcal{N}=1$ SU(3) gauge theory with 10 flavors whose spiral structure continues infinitely. (b) Its building block. 
}
\label{fig:tao}
\end{figure}

\begin{figure}
\includegraphics[width=15cm]{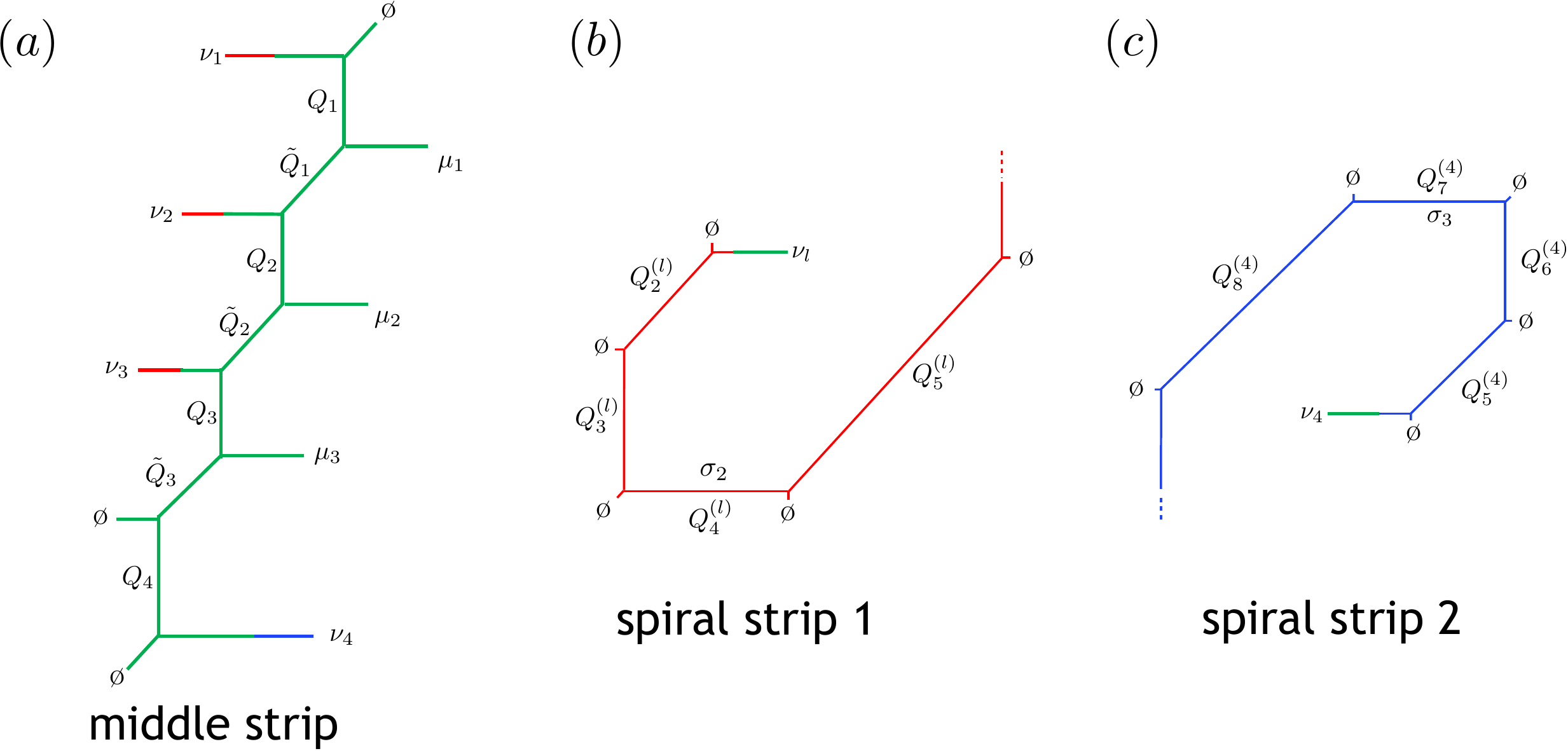}
\caption{The three kinds of building blocks of Figure \ref{fig:tao} (b). Here we call them (a) middle strip, (b) spiral strip 1, and (c) spiral strip 2. Greek letters $\nu_l \, (i=1,2,3)$ denote the Young diagrams.}
\label{fig:tao-strip}
\end{figure}

We first briefly review the computation of the partition function for 5d $\mathcal{N}=1$ SU(3)$_0$ gauge theory with 10 flavors based on a 5-brane web \cite{Hayashi:2016abm}. A 5-brane configuration for such 5d marginal theories is called Tao web diagram \cite{Kim:2015jba}, which is of a spiral shape with a periodic structure whose periodicity is given by the instanton factor squared. A Tao web diagram for 5d $\mathcal{N}=1$ SU(3)$_0$ gauge theory with 10 flavors is depicted in Figure \ref{fig:tao}(a). 
Though this 5-brane configuration, in principle, has infinitely many K\"ahler parameters, they are not independent due to the periodic structure 
of the web diagram, and only 13 parameters corresponding to the two Coulomb moduli parameters, ten mass parameters, and one instanton factor are independent.
For computations, only half of the diagram suffices due to the symmetry of the web diagram. For instance, in order to compute Figure \ref{fig:tao}(a), we only need to consider Figure \ref{fig:tao}(b). This diagram in Figure \ref{fig:tao}(b) can be further untangled with basic building blocks painted in different colors: middle strips, spiral strip 1, and spiral strip 2 as in Figure \ref{fig:tao-strip}. The assignment of the independent K\"ahler parameters ($Q_i$) and Young diagrams ($\mu_I$) are depicted in Figure \ref{fig:tao-strip}. The (unrefined) partition function is then obtained by gluing these building blocks with suitable edge factors.

The partition function for the 5d SU(3)$_0$ gauge theory with 10 flavors then reads  
\begin{align}
Z^{{\rm SU(3)+10F}}= \sum_{\mu_{1,2,3}} Z_{{\rm glue}} (Q_{b_1}, \mu_{1}) Z_{{\rm glue}} (Q_{b_2}, \mu_{2}) Z_{{\rm glue}} (Q_{b_3}, \mu_{3}) Z_{{\rm half \, 1}} Z_{{\rm half \, 2}}.
\end{align}
Here, $Z_{{\rm glue}}$ are the three edge factors for gluing, given by
\begin{align}
Z_{{\rm glue}}(Q,\mu) 
&= (-Q)^{|\mu|} q^{\frac{||\mu||^2+||\mu^2||^2}{2}} \prod_{(i,j)\in\mu} \frac{1}{(1-q^{\mu_i+\mu_j^t-i-j+1})^{2}},
\end{align}
where $q=e^{-\beta\epsilon}$ is the parameter associated with the string coupling, defined with the selfdual $\Omega$-deformation parameter $\epsilon= \epsilon_1 = -\epsilon_2$.
$Z_{{\rm half \, 1}}$ and $Z_{{\rm half \, 2}}$ are the contributions of the half diagram in Figure \ref{fig:tao}(b) and the other half, which are composed of three building blocks in Figure \ref{fig:tao-strip}, 
\begin{align}
Z_{{\rm half \, 1}}
&=  \sum_{\nu_{1,2,3,4}} Z_{{\rm middle}}(\nu_1,\nu_2,\nu_3,{\Emptyset} 
,\mu_1,\mu_2,\mu_3,\nu_4,\{ {\bm Q} \},\{ \tilde{{\bm Q}} \})
\nonumber \\
&\qquad \times \prod_{l=1}^3 Z_{\rm{spiral \, 1}}(\{ {\bm Q}^{(l)} \},\nu_l) \times Z_{\rm{spiral \, 2}} (\{ {\bm Q}^{(4)}\},\nu_4),
\label{eq:Z_half1}
\end{align}
where
\begin{subequations}
\begin{align}
&\{ {\bm Q} \} = \{ Q_1, Q_2, Q_3,Q_4\},\quad
 \{ \tilde{{\bm Q}} \} = \{ \tilde{Q}_1, \tilde{Q}_2, \tilde{Q}_3\}
 \\
&\{ {\bm Q}^{(k)} \} =  \{ Q_1^{(k)} , Q_2^{(k)}, Q_3^{(k)}, ... \} ~(k=1,2,3,4)
\end{align}
\end{subequations}
Each of which takes the form\footnote{\eqref{eq:Zmiddle} is slightly different from Eq. (3.8) of \cite{Hayashi:2016abm}, which has a typo in it. This typo in version 2 of \cite{Hayashi:2016abm} is corrected in \eqref{eq:Zmiddle}.
}
\begin{subequations}
\begin{align}
&Z_{{\rm middle}}(\mu_1,\mu_2,\mu_3,\mu_4,\nu_1,\nu_2,\nu_3,\nu_4,\{ {\bm Q}, \tilde{{\bm Q}} \})
\nonumber\\ \label{eq:Zmiddle}
&\qquad
= \frac{
\displaystyle 
\prod_{1\leq i \leq j \leq 4}R_{\mu_i \nu_j}\left(Q_j \prod_{k=1}^{j-1} Q_k \tilde{Q}_k \right)\prod_{1\leq i < j \leq 4}R_{\nu_i \mu_j}\left( \tilde{Q}_i \prod_{k=i+1}^{j-1} Q_k \tilde{Q}_k \right)
}{
\displaystyle 
\prod_{1\leq i \leq j \leq 4}R_{\mu_i \mu_j}\left(\prod_{k=1}^{j-1} Q_k \tilde{Q}_k \right)\prod_{1\leq i < j \leq 4}R_{\nu_i \nu_j}\left(\prod_{k=i}^{j-1} Q_{k+1} \tilde{Q}_k \right)
},
\\
&Z_{\rm{spiral \, 1}}(\{ {\bm Q} \},\nu_l) 
= \sum_{\sigma_{2,3,..}}
 Z_{{\rm glue}}(Q_{1},\nu_l)  \frac{R_{\nu_l \phi}(Q_{2}) R_{\phi \sigma_{2}^t}(Q_{3})}{R_{\nu_l \sigma_{2}^t}(Q_{2}Q_{3})}
\nonumber \\
&\hspace{38mm}  \times 
 \prod_{k=2}^\infty Z_{{\rm glue}}(Q_{3k-2},\sigma_k)  \frac{R_{\sigma_k \phi}(Q_{3k-1}) R_{\phi \sigma_{k+1}^t}(Q_{3k})}{R_{\sigma_k \sigma_{k+1}^t}(Q_{3k-1}Q_{3k})},
\\
&Z_{\rm{spiral \, 2}}(\{ {\bm Q}^{(4)}\},\nu_4)
= \sum_{\sigma_{3,4,...}} 
Z_{{\rm glue}}(Q_{4},\nu_4) \frac{R_{\nu_4 \phi}(Q_{5}) R_{\phi \sigma_{3}^t}(Q_{6})}{R_{\nu_4 \sigma_{3}^t}(Q_{5} Q_{6})}
\nonumber \\
&\hspace{38mm} \times
 \prod_{k=3}^\infty Z_{{\rm glue}}(Q_{3k-2},\sigma_k) \frac{R_{\sigma_k \phi}(Q_{3k-1}) R_{\phi \sigma_{k+1}^t}(Q_{3k})}{R_{\sigma_k \sigma_{k+1}^t}(Q_{3k-1} Q_{3k})}.
\end{align}
\end{subequations}
Here we define some quantities:
$\nu_i$ are the Young diagrams along the horizontal lines.
$R_{\mu \nu}(Q)$ is a function giving the contributions coming from the string wrapping on the internal line characterized by K\"ahler moduli $Q$ with Young diagrams $\mu$ and $\nu$,
\begin{align}
R_{\mu\nu}(Q) = \prod_{i,j=1}^\infty (1-Q q^{i+j-\mu_i-\nu^t_j -1}).
\end{align}
$Z_{{\rm half2}}$ can be obtained by suitable parameter replacement as the lower half (half1) and upper half (half2) of the diagram are symmetric.

After removing the extra factor, and using the proper flop transition \eqref{flop} for the perturbative part, one finds that
\begin{align}
&Z^{{\rm SU(3)+10F}} ={\rm PE} \left[F_0(\{ {\bm A} \}, \{ {\bm M} \}, q) + F_1(\{ {\bm A} \}, \{ {\bm M} \}, q) \mathfrak{q} + F_2(\{ {\bm A} \}, \{ {\bm M} \}, q) \mathfrak{q}^2 + \mathcal{O}(\mathfrak{q}^3)  \right],
\end{align}
where $\{ {\bm A} \} = \{ A_1, A_2, A_3 \}$ and $ \{ {\bm M} \} = \{ M_1, M_2,..., M_{10} \}$ are Coulomb branch moduli and mass parameters, respectively.
We just provide the explicit map between K\"ahler parameters and physical parameters in Appendix \ref{app:Kahler}. 
A detailed derivation is given in \cite{Hayashi:2016abm}.
We also define $F_{0,1,2}(A,M_i)$ as 
\begin{subequations}
\begin{align} 
&F_0 (\{ {\bm A} \}, \{ {\bm M} \}, q)= \frac{q}{(1-q)^2} \left[ \sum_{I=1}^3 \left( -\sum_{i=1}^{10}\frac{A_I}{M_i} + \sum_{J=1}^3 \frac{A_I}{A_J} \right) \right],\label{eq:F0}
\\
&F_1 (\{ {\bm A} \}, \{ {\bm M} \}, q)=\frac{q}{(1-q)^2} \Biggl[ \left(  \frac{\tilde{\chi}_0 A_1^4 \prod_{i=1}^{10}(1-A_1^{-1} M_i)}{ (1-A_1^{-1}A_2)^2 (1 -A_1^{-1} A_3)^2} + (\text{cyclic})
 \right) 
\nonumber \\
&\hspace{30mm} - \tilde{\chi}_1\left(A_1 + A_2 + A_3 \right)^2 - \tilde\chi_9 \left(A_1^{-1} + A_2^{-1} + A_3^{-1} \right)^2 + \tilde{\chi}_2 + \tilde{\chi}_8
\Biggr],\label{eq:F1}
\\
&F_2 (\{ {\bm A} \}, \{ {\bm M} \}, q) )=G_2 (\{ {\bm A} \}, \{ {\bm M} \}, q) - \frac{1}{2} F_1^2 (\{ {\bm A} \}, \{ {\bm M} \}, q) +\frac{1}{2} F_1 (\{ {\bm A}^2 \}, \{ {\bm M}^2 \}, q^2),
\label{eq:F2}
\end{align}
\end{subequations}
where
\begin{align}
G_2(\{ {\bm A} \}, \{ {\bm M} \}, q)  
= ~&\frac{q^2 }{(1-q)^4 }  \sum_{m=0}^{10} \sum_{n=0}^{10} (-1)^{m+n} 
\tilde{\chi}_m \tilde{\chi}_n 
\cr
&
\Biggl[
\frac{
A_1{}^{6-m} A_2{}^{6-n}
}{
 (A_1-A_3)^2 (A_2-A_3)^2 (A_1 - A_2 q^{-1})^2 (A_1-A_2 q)^2}
\cr
& 
~+ \frac{q^{8-n}  A_1{}^{12-m-n} 
}{
(1+q)^2 (A_1-A_2)^2 (A_1-A_3)^2 (A_2 - A_1 q)^2 (A_3-A_1 q)^2}
\cr
&
~+ \frac{q^{-6+n} A_1{}^{12-m-n} 
}{
(1+q)^2 (A_1-A_2)^2 (A_1-A_3)^2 (A_2 - A_1 q^{-1})^2 (A_3-A_1 q^{-1})^2}
\Biggr]
\cr
& 
+ ({\rm cyclic}), 
\end{align}
and
\begin{align}
\tilde\chi_n = \tilde\chi_0\,
 \sum_{1 \le i_1 < i_2 < \cdots < i_n \le10} M_{i_1} M_{i_2} \cdots  M_{i_n} 
\quad (n=1,2,\cdots 10), \qquad \tilde\chi_{0} = \prod_{i=1}^{10} M_i^{-\frac{1}{2}}. 
\end{align}
A letter ``(cyclic)" means two more terms that are obtained by taking a cyclic permutation of $A_I~(I=1,2,3)$ on the first term.
$\{ {\bm A}^2 \}$ and $\{ {\bm M}^2 \}$ denote the squares of $A_I$ and $M_i$,
\begin{align}
\{ {\bm A}^2 \} = \{ A_1^2, A_2^2, A_3^2 \},\quad
\{ {\bm M}^2 \} = \{ M_1^2, M_2^2,...,M_{10}^2 \}.
\end{align}

\subsection{5-brane web with two O5-planes}\label{sec;equivO5}
\begin{figure}
\centering
\includegraphics[width=12cm]{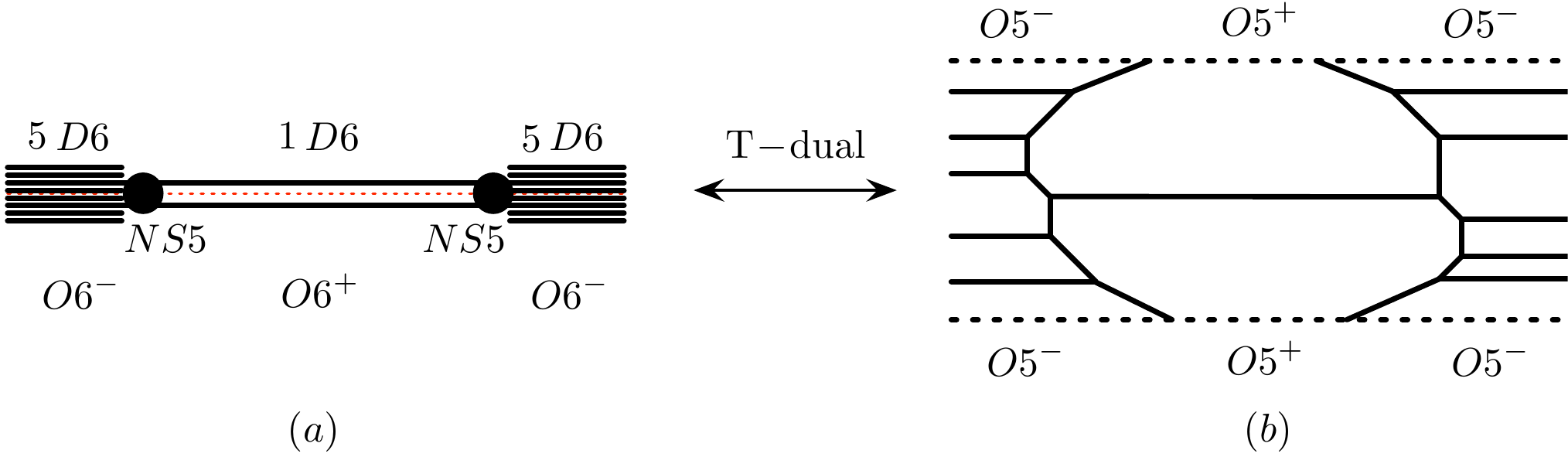}
\caption{(a): A Type IIA brane configuration for 6d Sp(1) gauge theory with 10 flavors and a tensor. (b): A Type IIB brane configuration which is T-dual of (a).}
\label{fig:T-dual}
\end{figure}
\begin{figure}[htb]
\centering
\includegraphics[width=15cm]{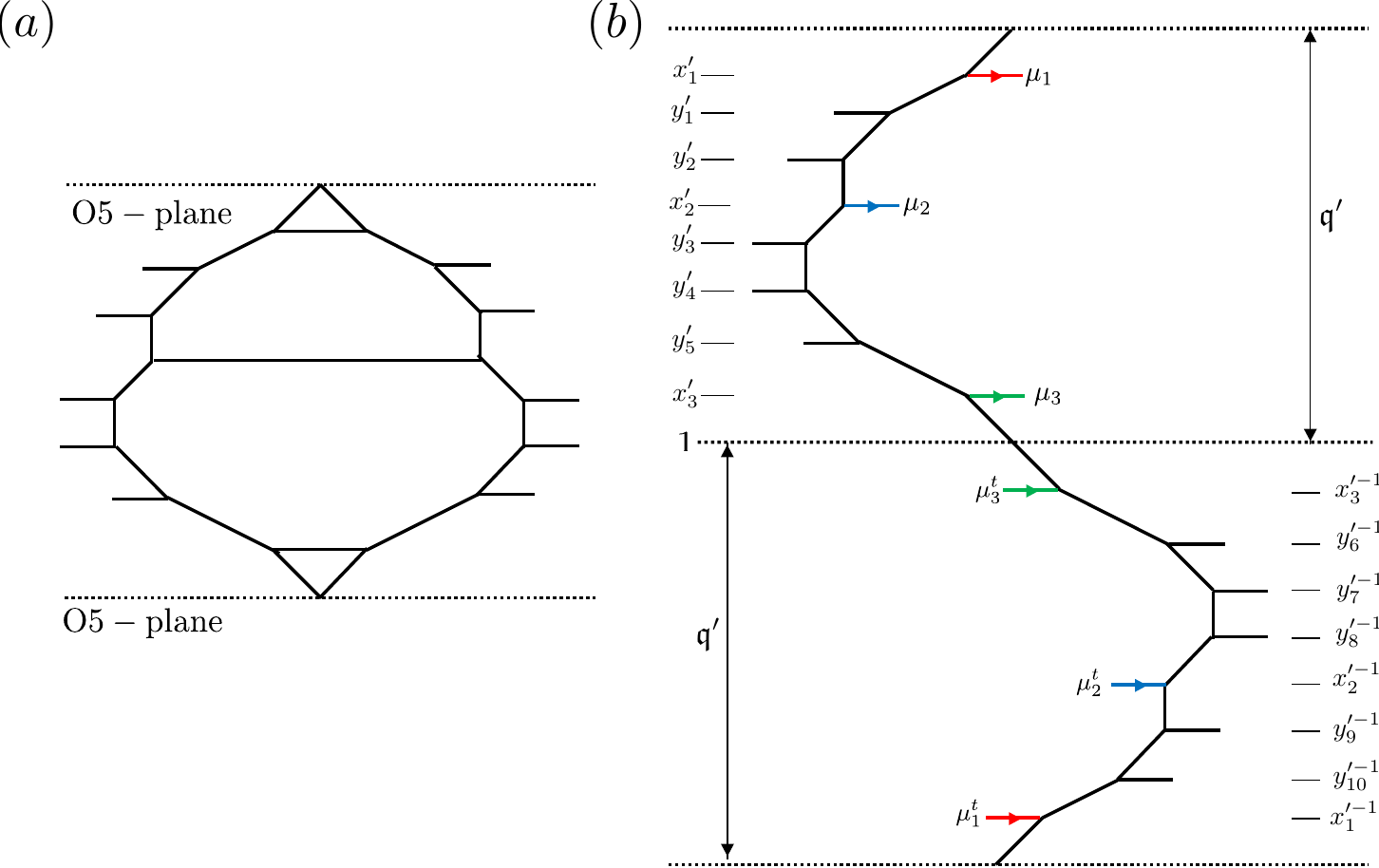}
\caption{A 5-brane web diagram description of 5d $\mathcal{N}=1$ Sp$(2)$ gauge theory with 10 flavors. 
The dashed lines and the solid lines denote the O$5$-planes and $(p,q)$ 5-branes, respectively.
The variables $x'_I~(I=1,2,3)$ and $y'_i~(i=1,2,...,10)$ denote the position of the horizontal legs measured from the middle of the dashed line, and $\mathfrak{q}'$ denotes the instanton factor.
}
\label{webO5}
\end{figure}
We now compute the partition function based on a 5-brane web with two O5-planes. 
The corresponding web describes 5d Sp(2) gauge theory with 10 flavors. As it is UV-dual to 6d Sp(1) gauge theory with 10 flavors and a tensor multiplet, a brane configuration for 5d Sp(2) gauge theory with 10 flavors can be obtained as a T-dual version of a Type IIA brane configuration for the 6d theory, which is made out of D6-branes, an O6-plane, and NS5 brane as given in Figure \ref{fig:T-dual}(a) \cite{Brunner:1997gf, Hanany:1997gh}. 
The corresponding Type IIB 5-brane configuration for 5d Sp(2) gauge theory with 10 flavors is a 5-brane configuration with two O5-planes as given by Figure \ref{fig:T-dual}(b).
As discussed in~\cite{Hayashi:2017btw}, a 5-brane web diagram in Figure \ref{fig:T-dual}(b) can be deformed into a 5-brane web diagram given in Figure~\ref{webO5}(a). 
Such deformed diagram gives rise to a strip-like diagram when the fundamental configuration is chosen to combine a half from the original diagram and the other from its reflected image due to the O5-plane, as depicted in Figure \ref{webO5}(b), where the top of $(1,1)$-brane is gluing to the bottom of $(-1,-1)$-brane by flipping the sign of the charges, i.e. the web diagram is compactified with the period ${\mathfrak{q}'}^2$.
This is a crucial 5-configuration which enables one to implement the topological vertex method to a 5-brane web with O5-plane(s) \cite{Kim:2017jqn}.
The computation in fact can be done in a straightforward way. With the fundamental configuration as a strip-like diagram given in Figure~\ref{webO5}(b), one performs the topological vertex computation with the following additional procedures: (i) the framing factor associated with a 5-brane crossing an O5-plane needs to be shifted by 1. (ii) internal edges of the web diagram associated with color D5-branes are glued in the following way: as  the configuration on the right hand side of Figure \ref{webO5}(b) are reflected, one takes the transpose for the Young diagrams assigned to the internal edges, and then one glues these internal edges together.  See Appendix \ref{TVFO5} for more details.

Based on the 5-brane configuration in Figure \ref{webO5}(b), the partition function for 5d Sp(2) gauge theory with 10 flavors is expressed in terms of a Young diagram sum over $\mu_1, \mu_2, \mu_3$, as  
\begin{align}
{Z'}^{{\rm Sp(2)+10F}}
&=
\sum_{\mu_{1,2,3}} 
\left( \frac{{\mathfrak{q}'}^2 x'_2 x'_3}{{x'}_1^4} \right)^{|\mu_1|}
\left(\frac{\prod_{i=1}^{10}y'_i}{{x'}_1 {x'}_2^6 x'_3} \right)^{|\mu_2|} 
\left(\frac{\prod_{i=1}^{10}y'_i}{{x'}_1 x'_2 {x'}_3^6} \right)^{|\mu_3|} 
f_{\mu_1}^{-5}f_{\mu_2}^{-5}f_{\mu_3}^{-5}
\nonumber\\
&\qquad\qquad\times \prod_{I=1}^3 \left(\prod_{i=1}^{10} \frac{\Theta_{\mu_I {\Emptyset}}(x'_I {y'}_i^{-1})}{\Theta_{\mu_I {\Emptyset}}(x'_I y'_i)} \prod_{J=1}^3 \frac{\Theta_{\mu_I \mu_J}(x'_I x'_J)}{\Theta_{\mu_I \mu_J^t}(x'_I {x'}_J^{-1})} \right),
\label{Sp210F}
\end{align}
where $f_\mu$ is the framing factor defined in \eqref{framingfactor}.
The positions $x'_I$ and $y_i$ correspond to the fugacities associated with the color D5-branes and those for 10 flavor masses, respectively.
For convenience, we have used $\Theta_{\mu\nu}(Q)$ defined as
\begin{align}\label{eq:Theta}
\Theta_{\mu\nu}(Q) = \prod_{n=0}^\infty (1-Q {\mathfrak{q}'}^{2n} q^{i+j-\mu_i - \nu_j -1}) (1-Q^{-1} {\mathfrak{q}'}^{2(n+1)} q^{i+j-\mu_i^t - \nu_j^t -1}),
\end{align}
We note that $\Theta_{\mu\nu}(Q)$ can be expressed as the Jacobi theta functions \eqref{eq:JacobiTheta} by using the analytic continuation formula \cite{Haghighat:2013gba}, and hence it is of periodic structure.
Notice that there are three Young diagram sums associated with color D5-branes in \eqref{Sp210F}. As discussed earlier, 
the 5d $\mathcal{N}=1$ Sp$(N)$ gauge theory with $N_f$ flavors is dual to the 5d $\mathcal{N}=1$ SU$(N+1)_\kappa$ gauge theory with $N_f$ flavors and the Chern-Simons level $\kappa=N+3-N_f/2$, through some kind of geometric transition \cite{Gaiotto:2015una, Hayashi:2015vhy, Hayashi:2015zka, Hayashi:2016abm}.
This enables us to relate the  parameters of the Sp($N$) theory and those of the SU($N+1$) theory. In this case, with a shifting factor $ \mathfrak{q}^{1/2} \Lambda_{\text{SU(3)}}$ for the Coulomb branch where $\Lambda_{\text{SU(3)}}= \prod_{i=1}^{10} M_i^{-1/4}$, the parameter map between the Sp(2) gauge theory with 10 flavors ($x'_I, y'_i$) and the SU(3)$_0$ gauge theory with 10 flavors ($A_I, M_i$) is given by \cite{Hayashi:2016abm} 
\begin{subequations}\label{SpSU10F}
\begin{align}
&x'_I =  \mathfrak{q}^{1/2} A_{I} \Lambda_{\text{SU(3)}}~~(I=1,2,3),
\\
&
y'_i = \mathfrak{q}^{1/2} M_i \Lambda_{\text{SU(3)}}~~(i=1,\cdots,10),
\end{align}
\end{subequations}
where the SU(3) Coulomb branch moduli $A_I$ satisfy $A_{1} A_{2} A_{3}=1$. 
We note that the instanton factors for these two dual theories are same $\mathfrak{q}' = \mathfrak{q}$, as they are associated with the compactification radius.

With the SU(3) gauge theory parametrization, we find the partition function of SU(3)$_0$ gauge theory with 10 flavors in a symmetric form,
\begin{align}\label{eq:TopZSU310F}
{Z'}^{{\rm Sp(2)+10F}}
=~&
Z^{\rm extra}
\sum_{\mu_{1,2,3}} 
\left( \mathfrak{q} \Lambda_{\text{SU(3)}}^{-2} A_{1}^{-5} \right)^{|\mu_1|}
\left(\mathfrak{q} \Lambda_{\text{SU(3)}}^{-2} A_{2}^{-5} \right)^{|\mu_2|} 
\left(\mathfrak{q} \Lambda_{\text{SU(3)}}^{-2} A_{3}^{-5} \right)^{|\mu_3|} 
f_{\mu_1}^{-5}f_{\mu_2}^{-5}f_{\mu_3}^{-5}
\nonumber\\
&\times \prod_{I=1}^3 \left(\prod_{i=1}^{10} \frac{\Theta_{\mu_I {\Emptyset}}(A_{I} M_i^{-1})}{\Theta_{\mu_I {\Emptyset}}(\mathfrak{q} A_{I} M_i \Lambda_{\text{SU(3)}}^2)} \prod_{J=1}^3 \frac{\Theta_{\mu_I \mu_J}(\mathfrak{q} A_{I} A_{J} \Lambda_{\text{SU(3)}}^2 )}{\Theta_{\mu_I \mu_J^t}(A_{I} A_{J}^{-1})} \right),
\end{align}
where $Z^{\rm extra}$ is an Coulomb branch moduli independent part which is given by  
\begin{align}
Z^{\rm extra}
=\prod_{i,j=1}^\infty 
\left[
 \frac{1}{(q^{i+j-1}\mathfrak{q};\mathfrak{q})_\infty^{10}} \times \frac{\prod_{1\leq i , j\leq 5} \Theta_{{\Emptyset}{\Emptyset}}(\mathfrak{q} M_i M_{j+5} \Lambda_{\text{SU(3)}}^2)}{\prod_{1\leq i<j\leq 5} \Theta_{{\Emptyset}{\Emptyset}}(M_i M^{-1}_j)\Theta_{{\Emptyset}{\Emptyset}}(M_{i+5} M^{-1}_{j+5})}
 \right],
\end{align}
where $(a;\mathfrak{q})_\infty =\prod_{k=0}^\infty (1-a \mathfrak{q}^k)$ is so-called Pochhammer symbol.
There are in fact more Coulomb branch independent part in \eqref{eq:TopZSU310F}, as a whole we call them the extra factor. 
When obtaining the partition function, we mod out such an extra factor from the topological string partition function.  
From here on, we neglect the extra factor.

By expressing the partition function \eqref{eq:TopZSU310F} as an expansion of the instanton factor $\mathfrak{q}$, we can write the partition function for SU(3)$_0$ gauge theory with 10 flavors as the Plethystic exponential, it takes the following form 
\begin{align}
&{Z'}^{{\rm Sp(2)+10F}}={\rm PE} \left[F_0(\{ {\bm A} \},\{ {\bm M} \}) + F_1(\{ {\bm A} \},\{ {\bm M} \}) \mathfrak{q} + F_2(\{ {\bm A} \},\{ {\bm M} \}) \mathfrak{q}^2 + \mathcal{O}(\mathfrak{q}^3)  \right],
\end{align}
where $F_n(\{ {\bm A} \},\{ {\bm M} \})$ are exactly the same as those obtained from Tao diagram \eqref{eq:F0}, \eqref{eq:F1}, and \eqref{eq:F2} since they are the dual to each other, although the partition functions $Z^{{\rm SU(3)+10F}}$ and ${Z'}^{{\rm Sp(2)+10F}}$ are derived from completely different diagrams\footnote{As we have already mentioned, these partition functions are the same up to extra factor. }.
Due to computational complication, we only presented terms of quadratic order in $\mathfrak{q}$, but can be checked the equivalence to higher orders.

\bigskip

\bigskip

\section{Defect partition function of E-strings}\label{sec:estring}
In this section, we consider defect insertions to the E-string theory, from the 5-brane perspective.
Through the geometric transitions, we can introduce a codimension 2 defect which is a topological brane wrapping on the Lagrange submanifold in the topological string theory \cite{Dimofte:2010tz}.
We consider the partition function of the topological string in the presence of the topological brane, which we call, for short, the defect partition function.

To obtain the defect partition function\footnote{We note that one can also introduce the holonomy matrix for the 5-brane in the topological vertex formalism as has been done e.g., in \cite{Aganagic:2003db}.} for the E-string theory, we utilize geometric transition \cite{Gopakumar:1998ki}.  
The procedure is to set the K\"ahler parameters associated with Higgsing $Q$ to the $M$-th power of the exponential of the string coupling constant, $Q=q^M$, where $M$ is the number of the topological branes.
In the context of the gauge theory, when $M=0$, it becomes the usual Higgsing, as $Q=1$ reduces the number of Coulomb branch moduli. 
When $M> 0$, it corresponds to the defects in the E-strings. We call such a  procedure for introducing the defects a defect Higgsing.  
More concretely, when we consider the 5d $\mathcal{N}=1$ Sp$(N)$ gauge theory with $2N+6$ flavors for $N\geq2$, the defect Higgsing means
\begin{align}
{\rm Sp}(N) + (2N+6)\text{ flavors} ~~\to~~ {\rm Sp}(N-1) + (2N+4)\text{ flavors} +\text{defects}.
\end{align}
Or, from the dual theory point of view, 
SU($N+1$) gauge theory with the same number of the flavors, it is 
\begin{align}
{\rm SU}(N+1) + (2N+6)\text{ flavors}~~ \to~~ {\rm SU}(N) + (2N+4)\text{ flavors} +\text{defects}.
\end{align}
In the following, we choose $N=2$ but $M>0$ arbitrary, which is relevant for $M$ defects in the E-strings. We explicitly compute the defect partition function for the E-string based on 5-brane webs by implementing the defect Higgsings.


\subsection{Usual Higgsing on 5-brane webs}  
\label{sec:HiggsingVSDefHiggsing}
\begin{figure}[htb]
\centering
\includegraphics[width=15cm]{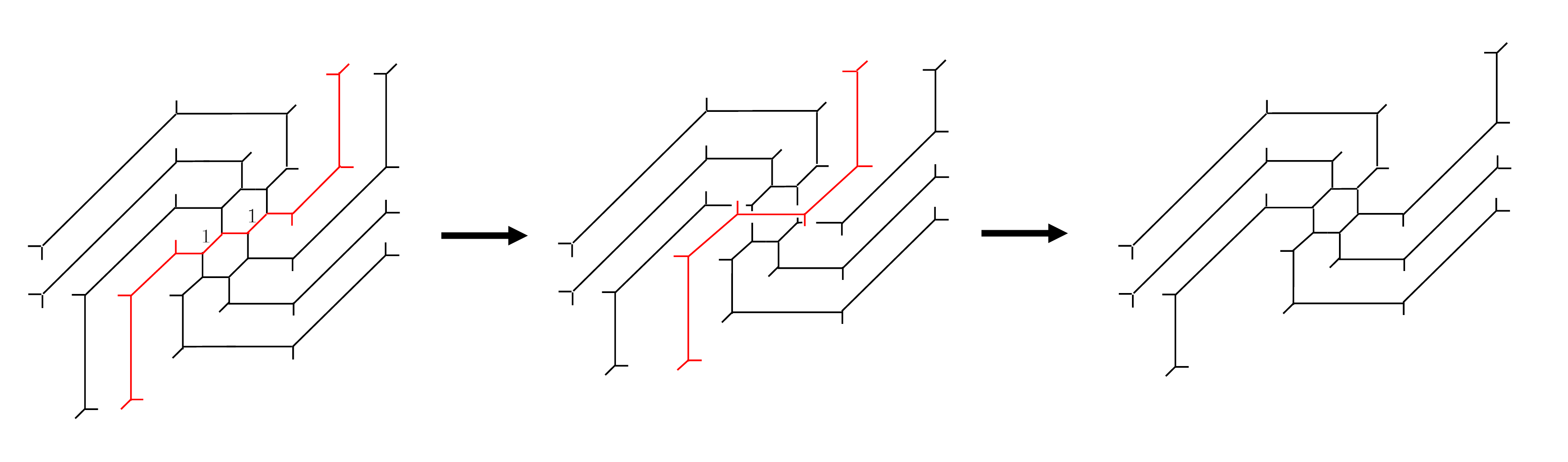}
\caption{Higgsing in a Tao diagram for SU(3) theory with 10 flavors. Since the framing denoted by the red lines decouples from the remaining web diagram, the part of the summation  of the Young diagram is decoupled from other summations.}
\label{taohiggs}
\end{figure}

Before we consider the defect Higgsing, we first review the usual Higgsing procedure in 5-brane configurations with or without O5-planes.
As done in the previous section, we consider 5d $\mathcal{N}=1$ SU(3) gauge theory with 10 flavors and then consider $\mathcal{N}=1$  Sp(2) gauge theory with 10 flavors.
We will show that one obtains SU(2) gauge theory with 8 flavors, as a result of the Higgsing on both theories. 
As discussed earlier, the 5-brane web for SU(3) gauge theory with 10 flavors is a Tao diagram given in Figure~\ref{fig:tao}(a).
To apply a Higgsing to the SU(2) theory with 8 flavors, we consider a strip consisting of a color D5-brane and two flavor branes that are connected.
Unlike the typical 5-brane web, for a Tao diagram, such a strip is a spiral strip as painted in red in Figure~\ref{taohiggs}(a).
A usual Higgsing is then realized by assigning the relevant K\"ahler parameters to 1, so that such a spiral string can be Higgsed away, reducing the dimension of the Coulomb branch by one and the number of flavors by two.
A diagrammatic procedure is depicted in Figure~\ref{taohiggs}. 

For implementing the Higgsing in the topological vertex calculation, as before, we can consider half of the diagram in Figure \ref{fig:tao}(b).
More specifically, in Figure \ref{fig:tao-strip}(a), say $\tilde{Q}_2$~ is the K\"ahler parameter associated with the Higgsing, then we set $\tilde{Q}_2=1$.  We expect that the part involving the summation over the Young diagram $\mu_2$ associated with the edges that are Higgsed is factorized in the partition function. In particular, the contribution involving the Young diagram $\mu_2$ in the building block $Z_{\text{middle}}$ in \eqref{eq:Zmiddle} becomes 
\begin{align}
\frac{
R_{\nu_1 \mu_2}(Q_1 \tilde{Q}_1 Q_2)R_{\nu_2\mu_2}(Q_2) R_{\nu_3 \mu_3}(Q_3)R_{\nu_3 \nu_4}(Q_3 \tilde{Q}_3 Q_4)  R_{\mu_2 \nu_3}(1) R_{\mu_2 \phi}(Q_3 \tilde{Q}_3)
}{
R_{\nu_1 \nu_3}(Q_1 \tilde{Q}_1 Q_2)R_{\nu_2 \nu_3}(Q_2)R_{\nu_3 \nu_3}(1)R_{\nu_3 \phi}(Q_3 \tilde{Q}_3)R_{\mu_2 \mu_3}(Q_3) R_{\mu_2 \nu_4}(Q_3\tilde{Q}_3 Q_4)
},
\label{factor}
\end{align}
which vanishes unless $\mu_2=\nu_3$ due to the factor $ R_{\mu_2 \nu_3}(1)$\footnote{
See also \cite{Cheng:2018wll} for refined case. },
\begin{align}
R_{\mu_2 \nu_3}(1) = \delta_{\mu_2, \nu_3} R_{\mu_2 \mu_2}(1),
\label{cond}
\end{align}
so that \eqref{factor} becomes 1 under the Higgsing.
By combining another half building blocks, the summation of $\mu_2$ is decoupled from other summations, as we expected,
\begin{align}
Z^{{\rm SU(3)+10F}}~~ \rightarrow ~~
Z^{{\rm Tao}}_{{\rm Framing}}  ~ Z^{{\rm SU(2)+8F}} ,
\label{PFtao}
\end{align}
where $Z^{{\rm SU(2)+8F}}$ is the partition function of SU$(2)$ gauge theory with 8 flavors 
\cite{Kim:2015jba},
\begin{align}
Z^{{\rm SU(2)+8F}}=
\sum_{\mu_{1,3},\{ \nu_i \}} Z_{{\rm glue}} (Q_{b_1}, \mu_1)  Z_{{\rm glue}} (Q_{b_3}, \mu_3) Z^{{\rm SU(2)+8F}}_{{\rm half \,  1}} Z^{{\rm SU(2)+8F}}_{{\rm half \,  2}},
\end{align}
with
\begin{subequations}
\begin{align}
& Z^{{\rm SU(2)+8F}}_{{\rm half \, 1}} =  \sum_{\nu_{1,2,4}} Z^{{\rm SU(2)}}_{{\rm middle}}(\nu_1,\nu_2, {\Emptyset}, 
\mu_1,\mu_3,\nu_4,\{ {\bm Q} \},\{ \tilde{{\bm  Q}} \})
\nonumber \\
&\qquad\qquad\qquad \times \prod_{l=1,2} Z_{\rm{spiral \,  1}}(\{ {\bm Q^{(l)}} \},\nu_l) \times Z_{\rm{spiral \,  2}} (\{ {\bm Q^{(4)}} \},\nu_4),
\\
&Z^{{\rm SU(2)+8F}}_{{\rm middle}}(\mu_1,\mu_3,\mu_4,\nu_1,\nu_2,\nu_4,\{ {\bm Q} \}, \{  \tilde{{\bm Q}} \} \})
\nonumber\\
&\qquad
=\frac{
\displaystyle 
\prod_{\substack{1\leq i \leq j \leq 4 \\ (i\neq2,j\neq3)}}R_{\mu_i \nu_j}\left(Q_j \prod_{k=1}^{j-1} Q_k \tilde{Q}_k \right)\prod_{\substack{1\leq i < j \leq 4 \\ (i\neq3,j\neq2)}}R_{\nu_i \mu_j}\left( \tilde{Q}_i \prod_{k=i+1}^{j-1} Q_k \tilde{Q}_k \right)
}{
\displaystyle 
\prod_{\substack{1\leq i \leq j \leq 4 \\ (i,j\neq2)}}R_{\mu_i \mu_j}\left(\prod_{k=1}^{j-1} Q_k \tilde{Q}_k \right)\prod_{\substack{1\leq i < j \leq 4 \\ (i,j\neq3)}}R_{\nu_i \nu_j}\left(\prod_{k=i}^{j-1} Q_{k+1} \tilde{Q}_k \right)
},
\\
&Z^{{\rm SU(2)+8F}}_{{\rm half \,  2}}  = Z^{{\rm SU(2)+8F}}_{{\rm half \,  1}} (M_i \leftrightarrow M^{-1}_{i+5};A_1 \leftrightarrow A_{3}^{-1}),
\end{align}
\label{BBSU2Tao}
\end{subequations}
and $Z^{{\rm Tao}}_{{\rm Framing}}$ can be interpreted as the contribution coming from the framing denoted by the red-colored strips  in Figure~\ref{taohiggs}, which is given by
\begin{align}
Z^{{\rm Tao}}_{{\rm Framing}}  = \sum_{\mu_{2}} Z_{{\rm glue}} (Q_{b_2}, \mu_2)  Z_{\rm{spiral \, 1}}(\{ {\bm Q}^{(3)} \},\mu_2)  R_{\mu_2 \mu_2}(1).
\end{align}

\begin{figure}
\centering
\includegraphics[width=15cm]{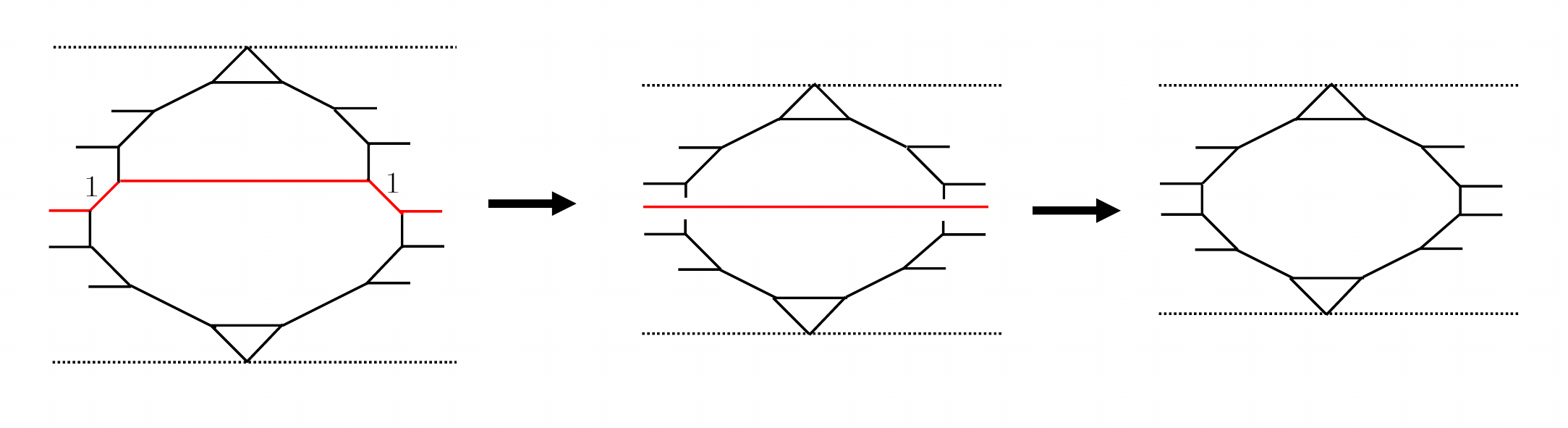}
\caption{
A web diagram description of a usual Higgsing. By setting the K\"ahler parameters to 1, one can  equate the Coulomb branch parameter and the fundamental masses. The D5-brane in red then can be taken to the Higgs branch directions. The resulting web diagram gives 5d $\mathcal{N}=1$ Sp$(1)$ gauge theory with 8 flavors.}
\label{webO5higgs}
\end{figure}
Now we consider the Higgsing from the 5-brane web in Figure \ref{webO5higgs} for the 5d $\mathcal{N}=1$ Sp(2) gauge theory with 10 flavors.\footnote{This Higgsing is also considered in \cite{Kim:2015fxa} in the context of the elliptic genus.}  
The Higgsing procedure is almost the same as that on the Tao diagram. We align a color brane and two flavor branes, for instance, 
\begin{align}
A_{2} = M_3=M_8 \, . 
\label{Higgs}
\end{align} 
It is also straightforward to see that the partition function \eqref{Sp210F} vanishes unless the associated Young diagram is  $\mu_2 = {\Emptyset}$, and hence it yields that the partition function  \eqref{Sp210F} reduces to 
\begin{align}
{Z'}^{{\rm Sp(2)+10F}}
~~ \rightarrow~~
Z^{{\rm O5}}_{{\rm Framing}}  {Z'}^{{\rm Sp(1)+8F}}.
\end{align}
Here $ {Z'}^{{\rm Sp(1)+8F}}$ is the partition function for the 5d $\mathcal{N}=1$ Sp(1) gauge theory with 8 flavors, given by
\begin{align}
& {Z'}^{{\rm Sp(1)+8F}}=\sum_{\mu_{1,3}} \left( \mathfrak{q} \Lambda^{-2}_{{\rm SU}(2)} A_1^{-4} \right)^{|\mu_1|}  \left( \mathfrak{q} \Lambda^{-2}_{{\rm SU}(2)} A_3^{-4} \right)^{|\mu_3|}  f_{\mu_1}^{-4} f_{\mu_3}^{-4} Z^{{\rm Sp(1)+8F}}_{{\rm build}},
\label{PFO5Higgs}
\end{align}
where
\begin{align}\label{BBSU2O5}
& Z^{{\rm Sp(1)+8F}}_{{\rm build}}
:= 
\prod_{I=1,3} \left(\prod_{i\in \mathcal{I}} \frac{\Theta_{\mu_I {\Emptyset}}(A_I M_i^{-1})}{\Theta_{\mu_I {\Emptyset}}(\mathfrak{q} A_I M_i \Lambda^2_{{\rm SU}(2)})} \prod_{J=1,3} \frac{\Theta_{\mu_I \mu_J}(\mathfrak{q} A_I A_J \Lambda^2_{{\rm SU}(2)} )}{\Theta_{\mu_I \mu_J^t}(A_I A_J^{-1})} \right)
\end{align}
with $\Lambda_{{\rm SU}(2)}=\prod_{i\in \mathcal{I}} M_i^{-1/4}$ and $\mathcal{I}=\{1,2,4,5,6,7,9,10 \}$.
The factor $Z^{{\rm O5}}_{{\rm Framing}}$ is an extra factor which comes from the contribution of the framing denoted by the red line in Figure \ref{webO5higgs},
\begin{align}
Z^{{\rm O5}}_{{\rm Framing}}=\prod_{i,j=1}^\infty (1-q^{i+j-1}).
\end{align}
By rescaling the Coulomb moduli parameters and the mass parameters as,
\begin{align}
A_1 \to M_3^{-1/2} A_1, \quad
A_3 \to M_3^{-1/2} A_3, \quad
M_i \to M_3^{-1/2} M_i~ (i\neq3,8),
\end{align}
\vspace{0.2cm}
we see that \eqref{PFO5Higgs} agrees with the partition function directly calculated from the web diagram of 5d $\mathcal{N}=1$ SU(2) with 8 flavors discussed in \cite{Kim:2017jqn}.

\begin{figure}
\centering
\includegraphics[width=15cm]{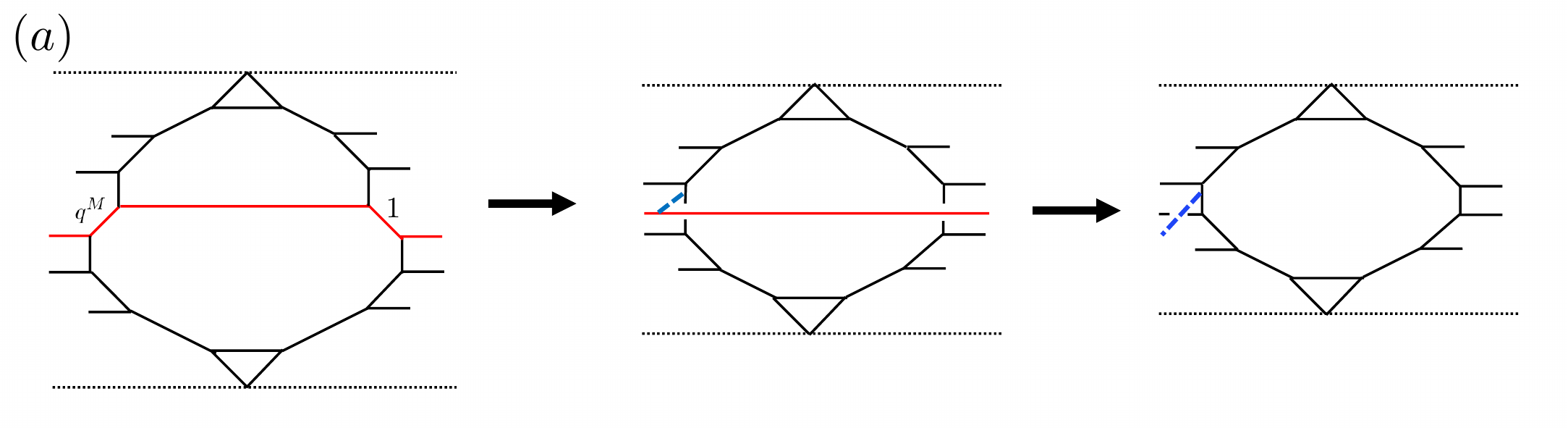}
\vspace{5mm}
\includegraphics[width=15cm]{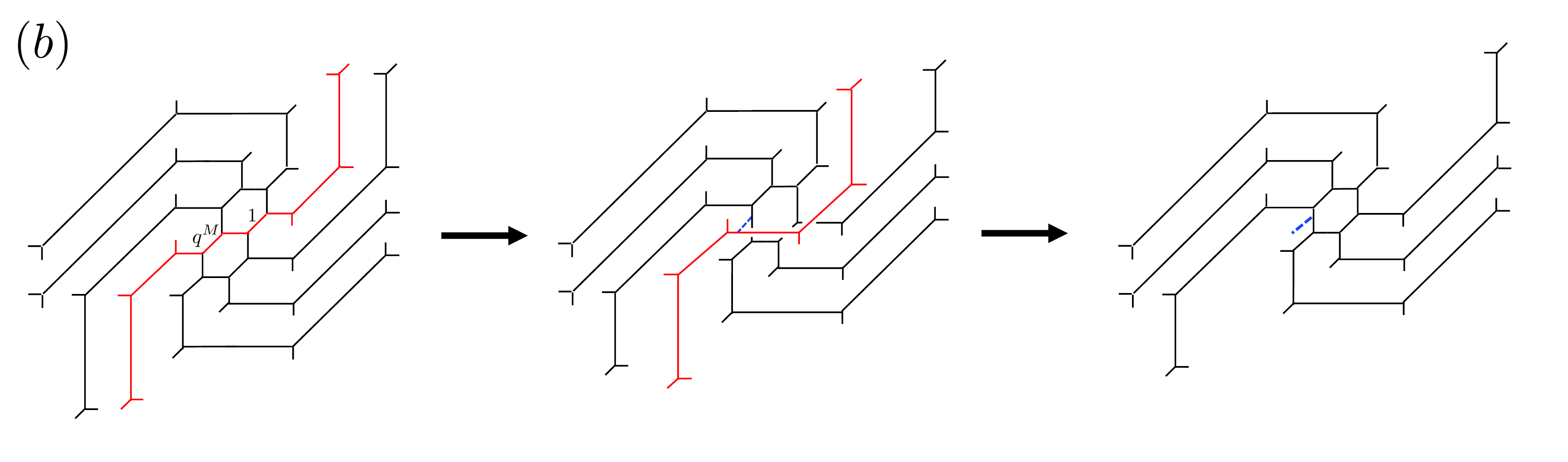}
\caption{The 5-branes with the defect introduced by defect Higgsing. The blue dashed lines denote the defects.}
\label{webdef}
\end{figure}
\subsection{Defect Higgsing on 5-brane webs}\label{sec:DefctHiggssingOn5-braneWebs}  
From here on, we consider a 5-brane system with defect D3-branes, which can be introduced in a similar way as for usual Higgsings. 
As discussed, the defect Higgsing is the Higgsing by setting the associated K\"ahler parameter not 1 but $q^M$ \cite{Alday:2009fs, Dimofte:2010tz, Taki:2010bj}. 
Here, $M$ is the number of defect D3-branes, and hence the defect Higgsing becomes a usual Higgsing when $M=0$.
For instance, in Figure \ref{webdef}(a), 5-brane web for the 5d Sp(2) theory with 10 flavors can be used to describe the 5d Sp(1) theory with 8 flavors and defects, where the 5-branes painted in red are defect Higgsed.
The associated K\"ahler parameter is set to be $q^M$, while the other K\"ahler parameter is just treated as a usual Higgsing.
After having defect Higgsed, we denote the defect Higgsed part as a (blue) dotted line, representing $M$  D3-branes.
Likewise, such defect Higgsing can be performed in a Tao diagram in the same way, as depicted in Figure \ref{webdef}(b).

Implementing the defect Higgsings to the partition function is straightforward. 
For the partition functions that we obtained from topological vertex based on the web diagrams either with O5-planes or with the Tao diagram, in the previous section, we first  
set one of the Coulomb branch moduli $A_2$ to
\begin{align}
A_2=q^M z = M_8,
\label{defHiggs}
\end{align}
where we defined $z=M_3$ as a defect moduli. When $M=0$, it reduces to \eqref{Higgs}, as expected. 
As the computation is similar to the usual Higgsing, we summarize the results for both cases.

\paragraph{Defect Higgsing on $SU(3)$ 5-brane web.}
Let us first consider the defect Higgsing on 5-brane web for $\mathcal{N}=1$ SU(3) gauge theory with 10 flavors, given in Figure \ref{webdef}(b). 
For the partition function \eqref{PFtao} obtained from the Tao web diagram,  
the defect Higgsing on SU(3) gauge theory with 10 flavors yields
\begin{align}
Z^{{\rm SU(3)+10F}}
\rightarrow
& \sum_{\mu_{1,3},\{ \nu_i \}} Z_{{\rm glue}} (Q_{b_1}, \mu_1)  Z_{{\rm glue}} (Q_{b_3}, \mu_3) Z^{{\rm SU(2)+8F}}_{{\rm half \, 1}} Z^{{\rm SU(2)+8F}}_{{\rm half \, 2}}
\cr
&\times
Z_{{\rm glue}} (Q_{b_2}, \nu_3)  Z_{\rm{spiral \, 1}}(\{ {\bm Q^{(3)}} \},\nu_3) Z_{\rm{defect \, Tao}}
\cr
=:&~{Z}^{\rm{SU(2)+8F, \, defect}},
\label{PFtaodef}
\end{align}
where 
\begin{align}
Z_{\rm{defect \, Tao}}=\,&
\frac{
R_{\nu_1 \nu_3}(Q_1 \tilde{Q}_1 Q_2)R_{\nu_2\nu_3}(Q_2) R_{\nu_3 \nu_3}(q^M)  R_{\nu_3 \mu_3}(Q_3) 
}{
R_{\nu_1 \nu_3}(q^M Q_1 \tilde{Q}_1 Q_2)R_{\nu_2 \nu_3}(q^M Q_2)R_{\nu_3 \nu_3}(1)
}\nonumber\\
& \times \frac{
R_{\nu_3 \nu_4}(Q_3 \tilde{Q}_3 Q_4)  R_{\nu_3 {\Emptyset}}(q^M Q_3 \tilde{Q}_3) R_{\mu_1 \nu_3}(q^M \tilde{Q}_1 Q_2)
}{
R_{\nu_3 \mu_3}(q^M Q_3) R_{\nu_3 \nu_4}(q^M Q_3\tilde{Q}_3 Q_4)R_{\nu_3 {\Emptyset}}(Q_3 \tilde{Q}_3)R_{\mu_1 \nu_3}(\tilde{Q}_1 Q_2)
},
\end{align}
and we have rescaled some variables as follows,
\begin{align}
A_I \to q^{-M/2} z^{-1/2} A_I,~(I=1,3),
\end{align}
to obtain the traceless condition of SU(2), $A_1 A_3 = 1$.
This then naturally shifts other mass parameters
\begin{align}
M_i \to q^{-M/2} z^{-1/2} M_i,~(i\neq 3,8).	
\end{align}
\vspace{0.3cm}

\paragraph{Defect Higgsing on $Sp(2)$ 5-brane web.}
Next, we consider the defect Higgsing on 5-brane web for $\mathcal{N}=1$ Sp(2) gauge theory with 10 flavors, given in Figure \ref{webdef}(a). 
 For the partition function \eqref{eq:TopZSU310F} obtained from the 5-brane web with O5-planes, the defect Higgsing on the Sp(2) gauge theory with 10 flavors yields
\begin{align}
{Z'}^{{\rm Sp(2)+10F}} 
\rightarrow
&~\prod_{i=1}^4 \frac{\Theta_{{\Emptyset}{\Emptyset}}(q^M z M_i^{-1}) \Theta_{{\Emptyset}{\Emptyset}}(\mathfrak{q} z M_{i+4} \Lambda_{\text{SU(2)}}^2)}{\Theta_{{\Emptyset}{\Emptyset}}(z M_i^{-1}) \Theta_{{\Emptyset}{\Emptyset}}(\mathfrak{q}z q^M M_{i+4} \Lambda_{\text{SU(2)}}^2)}
\nonumber \\&\times
\sum_{\mu_{1,3}} 
\left( \mathfrak{q}q^{M/2}z \Lambda_{\text{SU(2)}}^{-2} A_{1}^{-4}  \right)^{|\mu_1|} \left(\mathfrak{q} q^{M/2}z \Lambda_{\text{SU(2)}}^{-2} A_{3}^{-4} \right)^{|\mu_3|}
f_{\mu_1}^{-4}f_{\mu_3}^{-4}
\nonumber \\
&\times
 {Z'}^{{\rm Sp(1)+8F}}_{{\rm build}}(\Lambda_{\text{SU(2)}} \to q^{M/4}\Lambda_{\text{SU(2)}})\, Z^{\rm{defect \, O5}}
\nonumber \\
 =:&~{Z'}^{\rm{Sp(1)+8F, \, defect}},
\label{PFO5def}
\end{align}
where  $ {Z'}^{{\rm Sp(1)+8F}}_{{\rm build}}$ is defined in \eqref{BBSU2O5} with the replacement $\Lambda_{\text{SU(2)}} \to q^{M/4}\Lambda_{\text{SU(2)}}$, and  
\begin{align}
&Z^{\rm{defect \, O5}}
\nonumber \\
&=
 \frac{\Theta_{\mu_1 {\Emptyset}}(q^{-M/2} z^{-3/2} A_{1} )\Theta_{\mu_3 {\Emptyset}}(q^{-M/2} z^{-3/2} A_{3} )\Theta_{\mu_1 {\Emptyset}}(\mathfrak{q} q^{M} z^{3/2} A_{1} \Lambda_{\text{SU(2)}}^2)\Theta_{\mu_3 {\Emptyset}}(\mathfrak{q} q^{M} z^{3/2}  A_{3} \Lambda_{\text{SU(2)}}^2)}{\Theta_{\mu_1 {\Emptyset}}(q^{-3M/2} z^{-3/2} A_{1})\Theta_{\mu_3 {\Emptyset}}(q^{-3M/2} z^{-3/2} A_{3})\Theta_{\mu_1 {\Emptyset}}(\mathfrak{q}  z^{3/2} A_{1} \Lambda_{\text{SU(2)}}^2)\Theta_{\mu_3 {\Emptyset}}(\mathfrak{q}  z^{3/2} A_{3} \Lambda_{\text{SU(2)}}^2)}.
\end{align}
We note that though two defect partition functions \eqref{PFtaodef} and \eqref{PFO5def} look different, it is straightforward to see that they are equivalent when we expand as $\mathfrak{q}$ and $A_1$.

When expanding the partition function as a series of the instanton factor and the defect moduli, the partition function is decomposed into two parts: One is the defect moduli independent part which hence depends only on the Coulomb moduli and mass parameters. We denote it by $F^{\text{defect}}_n (A) $. The other is the part that explicitly depends on the defect moduli, which we denote by $D^{\text{defect}}_n (z,A)$, up to an extra factor, 
\begin{align}
{Z'}^{\rm{Sp(1)+8F, \, defect}}= {\rm PE}\left[\sum_{n=0}^\infty \mathfrak{q}^n \Big(F^{\text{defect}}_n (A) + D^{\text{defect}}_n (z,A) \Big) \right],\label{eq:Def8}
\end{align}
where $A=A_{1}$. Up to 2-instanton orders, $F^{\text{defect}}_n (A)$ and $D^{\text{defect}}_n (x,A)$ take the following forms. 
In terms of the U(8) characters,  
\begin{subequations}
\begin{align}
&\chi_n = \chi_0 \sum_{\substack{1 \leq i_1 < i_2 < ...< i_n \leq 10 \\ i_ m\in \mathcal{I}}} M_{i_1} M_{i_2} ... M_{i_n},\quad n=1,2,...,8,
\\
&\chi_0 = \prod_{i\in \mathcal{I}} M_i^{-1/2},
\end{align}
\end{subequations}
they are given by, at $\mathfrak{q}^0$ order, 
\begin{subequations}
\begin{align}
F^{\text{defect}}_0 (A) = &~\frac{q}{(1-q)^2} \left(-A\left(\chi_1 \chi_8 +\chi_0 \chi_7 \right) + 2A^2 \right),
\\ 
D^{\text{defect}}_0(z,A) =&~ \frac{q(q^{M/2}-q^{-M/2})}{(1-q)^2} A \left( q^M z^{3/2} - q^{-M} z^{-3/2} \right),
\end{align}
\end{subequations}
and, at $\mathfrak{q}^1$ order, 
\begin{subequations}
\begin{align}
F^{\text{defect}}_1 (A) =&-\frac{q}{(1-q)^2} \left(q^{M/2} \left(\chi_1 + \chi_3 \right)+q^{-M/2} \left(\chi_5+\chi_7\right)  \right)A + \mathcal{O}(A^2),
\\
D^{\text{defect}}_1(z,A)= & ~\frac{q(q^{M/2}-q^{-M/2})}{(1-q)^2}\biggl[
 \left(q^{9M/2} z^{9/2} \chi_0 - q^{-9M/2} z^{-9/2} \chi_8  \right)  - \left(q^{3M} z^{3} \chi_1 - q^{-3M} z^{-3} \chi_7  \right)  
\nonumber \\
&+\left(q^{3M/2}z^{3/2}\chi_2 -q^{-3M/2}z^{-3/2}\chi_6  \right)+\mathcal{O}(z^{15/2})
 \biggr]A + \mathcal{O}(A^2),
\end{align}
\end{subequations}
and at $\mathfrak{q}^2$ order, 
\begin{subequations}
\begin{align}
F^{\text{defect}}_2 (A) 
=&-\frac{q}{(1-q)^2} \biggl[
 q^M\left( \chi_0 \chi_5 + \chi_0 \chi_7 + \chi_1 \chi_8 \right)  + q^{-M}\left(\chi_3 \chi_8 + \chi_0 \chi_5 + \chi_1 \chi_8 \right) 
 \nonumber \\
&
- 2(q+q^{-1}+5)\left(\chi_0 \chi_7 + \chi_1  \chi_8\right)
 - \left(\chi_1 \chi_6  + \chi_2 \chi_7 - 7 \left( \chi_0 \chi_7 + \chi_1 \chi_8 \right) \right)
 \biggr]A 
 \nonumber \\
&
+ \mathcal{O}(A^2),
 \\
D^{\text{defect}}_2(z,A)=&~  \frac{q(q^{M/2}-q^{-M/2})}{(1-q)^2}\biggl[
q^{8M} \chi_0^2 z^{15/2}-q^{-8M} \chi_0^2 z^{-15/2}
 \nonumber \\
&
-q^{13M/2}\chi_0 \chi_1 z^{6}+q^{-13M/2}\chi_7 \chi_8 z^{-6} +q^{5M} \chi_0 \chi_2 z^{9/2}-q^{-5M} \chi_6 \chi_8 z^{-9/2}
 \nonumber \\
&
- \left( q^{7M/2} \chi_0  \chi_3 + q^{5M/2}  \chi_0 \chi_7 \right) z^3 +\left( q^{-7M/2} \chi_5  \chi_8 + q^{-5M/2}  \chi_1 \chi_8 \right) z^{-3} 
 \nonumber \\
&
+\left( q^{2M}\left(-1+\chi_0 \chi_4 \right) + q^M\left(2q+2q^{-1}+2+\chi_0 \chi_6 + \chi_1 \chi_7 \right) \right) z^{3/2}
 \nonumber \\
&
-\left( q^{2M}\left(-1+\chi_8 \chi_4 \right) + q^M\left(2q+2q^{-1}+2+\chi_2 \chi_8 + \chi_1 \chi_7 \right) \right) z^{-3/2}
 \nonumber \\
&
+\mathcal{O}(z^{21/2})
 \biggr]A + \mathcal{O}(A^2).
\end{align}
\end{subequations}
Here we note that the global symmetry is seemingly broken through the defect Higgsing, as the defect partition function is expressed in terms of the U(8) characters. However,
recall that the usual Higgsing from the SU(3) gauge theory with 10 flavors to the SU(2) gauge theory with 8 flavors preserves an $E_8$ symmetry, and such  enhanced global symmetry can be seen suitably combining flavors masses and instanton fugacity such that they form the characters of the enhanced global symmetry.
To make the global symmetry of the theory in the presence of the defects manifest, we consider the elliptic genus.



\subsection{Defect Higgsing in elliptic genus and global symmetry}\label{elliptic} 
We compute the elliptic genus of E-strings with defect by utilizing the defect Higgsing.
To this end, we use the result of \cite{Kim:2014dza}, where the 6d Sp(1) gauge theory with 10 flavors and a tensor is computed.  
The elliptic genus for one-string $\tilde{Z}_{(1)}$  and that for two strings  $\tilde{Z}_{(2)}$ are given by 
\cite{Kim:2014dza, Kim:2015fxa, Hayashi:2016abm},
\begin{subequations}
\begin{align}
\tilde{Z}_{(1)}  =\,& \frac{1}{2} \frac{\eta^2}{\theta^2_1(q)} \sum_{I=1}^4 \frac{\eta^2}{\theta_I(\tilde{A})\theta_I(\tilde{A}^{-1})} \prod_{l=1}^{10}\frac{\theta_I (\tilde{y_l})}{\eta},
\\
\tilde{Z}_{(2)} =
 &\,\frac{\eta^8}{\theta^2_1(q)\theta^2_1(\tilde{A}^2q)\theta^2_1(\tilde{A}^2q^{-1})\theta^2_1(\tilde{A}^2)} \prod_{l=1}^{10} \frac{\theta_1(\tilde{y}_l \tilde{A})\theta_1(\tilde{y}_l \tilde{A}
^{-1})}{\eta^2}
\nonumber\\
&+ \frac{1}{2} \frac{\eta^4}{\theta^2_1(q)\theta^2_1(q^2)} \sum_{I=1}^4 \frac{\eta^4}{\theta^2_I(\tilde{A}q^{1/2})\theta^2_I(\tilde{A}q^{-1/2})} \prod_{l=1}^{10}\frac{\theta_I (\tilde{y_l}q^{1/2})\theta_I (\tilde{y_l}q^{-1/2})}{\eta^2}
\nonumber\\
&+ \frac{1}{4} \frac{\eta^4}{\theta^4_1(q)} \sum_{(I,J,K)\in S} \frac{(-1)^{\delta_{K,1}} \eta^4 \theta^2_I(1)}{\theta^2_I(q)\theta^2_J(\tilde{A})\theta^2_K(\tilde{A})} \prod_{l=1}^{10}\frac{\theta_J (\tilde{y_l})\theta_K (\tilde{y_l})}{\eta^2},
\end{align}
\end{subequations}
where in the last line, $S=\{ (2,2,1), (3,3,1), (4,4,1),(2,3,4), (3,4,2), (4,2,3) \}$ and $\delta_{I,J}$ is the Kronecker delta. 
The $\tilde{y}_i\,(i\!=\!1,\cdots,10)$, $\tilde{A}$, $\phi$, and $\tilde{\mathfrak{q}}$ denote the fugacities for an SO(20) flavor symmetry, Sp(1) gauge symmetry, counting the number of the self-dual strings, and instanton factor, respectively.
The explicit forms of the Jacobi's theta functions $\theta_I(x)$ are given in Appendix \ref{sec:appendix}. The partition function of the 6d Sp(1) gauge theory then takes the form
\begin{align}
&\tilde{Z}^{{\rm 6d \, Sp(1)}} = {\rm PE}\left[\tilde{F}_{(0)} + \tilde{F}_{(1)} \phi + \tilde{F}_{(2)} \phi^2 + \mathcal{O}(\phi^3) \right],
\end{align}
where
\begin{align}
\tilde{F}_{(1)} = \tilde{Z}_{(1)},\qquad \tilde{F}_{(2)} = \tilde{Z}_{(2)}-\frac{1}{2} \tilde{Z}_{(1)}^2 -\frac{1}{2}  \tilde{Z}_{(1)}(\ast\to\ast^2).
\end{align}
The overall factor $\tilde{F}_{(0)}$ comes from ``zero string" contribution explained in 
\cite{Hayashi:2016abm},
\begin{align}
\tilde{F}_{(0)} = \frac{q}{(1-q)^2} \left( 2\left(\tilde{A}^2 + \tilde{A}^{-2} \right) -\left( \tilde{A} + \tilde{A}^{-1} \right) \sum_{i=1}^{10} \left( \tilde{y}_i + \tilde{y}_i^{-1} \right) \right)\left(\frac{\tilde{\mathfrak{q}}}{1-\tilde{\mathfrak{q}}} + \frac{1}{2} \right).
\end{align}
We summarize the duality map between two theories among the 6d Sp(1) gauge theory with 10 flavors and a tensor (6d Sp(1)$+10{\bf F}+1{\bf T}$), 5d Sp(2) gauge theory with 10 flavors (5d Sp(2)$+10{\bf F}$), and 5d SU(3)$_0$ gauge theory with 10 flavors (5d SU(3)$_0+10{\bf F}$):
\begin{itemize}
\item The duality map between 
	6d Sp(1)$+10{\bf F}+1{\bf T}$ and 5d Sp(2)$+10{\bf F}$:
\begin{equation}
\begin{aligned}
\tilde{y}_i &= {y'}^{-1}_i ~(i=1,...,5),
\qquad &
\tilde{y}_i &= y'_i ~(i=6,...,9),
\qquad &
\tilde{y}_{10} &= {\mathfrak{q}'}^{-2} y'_{10},
\\
\phi &= {\mathfrak{q}'} {y'}_{10}^{-1} A'_1,\qquad\quad
\tilde{A} = A'_2,\qquad &
\tilde{\mathfrak{q}}&={\mathfrak{q}'}^2,
\label{SpSp10F}
\end{aligned}
\end{equation}
where $y'_i\,(i=1,\cdots,10)$, $A'_j\,(j=1,2)$, $\mathfrak{q}'$ are the flavor fugacities, the Coulomb branch moduli, and instanton factor, respectively.
\item The duality map between 6d Sp(1)$\,+10{\bf F}+1{\bf T}$  and 5d SU(3)$_0+10{\bf F}$, which comes from \eqref{SpSp10F} and \eqref{SpSU10F} with $\mathfrak{q}'=\mathfrak{q}$: \begin{equation}
\begin{aligned}
\tilde{y}_i &= \mathfrak{q}^{-1/2}  
 \Lambda_{\text{SU(3)}}^{-1}\,M^{-1}_i~(i=1,..,5),\quad
&
\tilde{y}_i &= \mathfrak{q}^{1/2} 
\Lambda_{\text{SU(3)}}\,M_i~(i=6,..,9), 
\\
\tilde{y}_{10} &= \mathfrak{q}^{-3/2}  
\Lambda_{\text{SU(3)}} \,M_{10},
&\phi &= \mathfrak{q}^{2}  M_{10}^{-1} A_1,
\\
\tilde{A} &= \mathfrak{q}^{1/2} 
\Lambda_{\text{SU(3)}} \,A_2,\qquad
&\tilde{\mathfrak{q}}&=\mathfrak{q}^2
\label{map}.
\end{aligned}
\end{equation}
\end{itemize}
Using the duality map above, one can readily check that two partition functions do agree with each other by a double expanding in terms of the 5d Coulomb parameter $A_1$ and the instanton fugacity $\mathfrak{q}$ \cite{Hayashi:2016abm}.

Some comments are in order.
First, this transformation is slightly different from the one given in \cite{Hayashi:2016abm}, however, the difference is just the convention, and our convention is more useful to consider the flavor decoupling limit that we will discuss in Appendix \ref{decoupling}, so that we adopt them.
Second, we check the agreement of the partition function and the elliptic genus by the double expansion as a power series of $A_1$ and $\mathfrak{q}$ up to second order through the duality map \eqref{map}.

The usual Higgsing from 6d Sp(1) gauge theory with 10 flavors and a tensor, to 6d E-string theory, is achieved by setting the Coulomb branch parameter $\tilde{A}$ and two mass parameters $\tilde y_3, \tilde y_8$ as 
\begin{align}
\tilde{A}=\tilde{y}^{-1}_3 = \tilde{y}_8,
\end{align}
which is equivalent to \eqref{Higgs} through the map \eqref{SpSp10F}. Then the elliptic genera of one and two strings become 
\begin{subequations}
\begin{align}
\tilde{Z}_{(1)}  \rightarrow& \frac{1}{2} \frac{\eta^2}{\theta^2_1(q)} \sum_{I=1}^4  \prod_{l\in \mathcal{I}} \frac{\theta_I (\tilde{y_l})}{\eta},
\\
\tilde{Z}_{(2)} \rightarrow
& \frac{1}{2} \frac{\eta^4}{\theta^2_1(q)\theta^2_1(q^2)} \sum_{I=1}^4  \prod_{l\in \mathcal{I}} \frac{\theta_I (\tilde{y_l}q^{1/2})\theta_I (\tilde{y_l}q^{-1/2})}{\eta^2}
\nonumber \\
&\hspace{20mm}+ \frac{1}{4} \frac{\eta^4}{\theta^4_1(q)} \sum_{(I,J,K)\in S} \frac{ \theta^2_I(1)}{\theta^2_I(q)}\prod_{l\in \mathcal{I}} \frac{\theta_J (\tilde{y_l})\theta_K (\tilde{y_l})}{\eta^2},
\end{align}
\end{subequations}
which agree with those for the E-string theory up to redefinitions of the parameters.  
For the elliptic genus of one string, the summation of the product of the theta functions can be expressed as $E_8$ theta function which is defined as the summation over the $E_8$ root lattice $\Gamma_8$,
\begin{align}
 \sum_{I=1}^4  \prod_{l\in \mathcal{I}} \frac{\theta_I (\tilde{y_l})}{\eta} = 2 \sum_{\vec{w}\in \Gamma_8} {\rm exp}\left[\pi {\rm i} \tau \vec{w} + 2\pi {\rm i} \vec{\mu} \cdot \vec{w}  \right],
\end{align}
where $\vec{w}=(w_1,\cdots,w_8),~\vec{\mu}=(\mu_1,\cdots,\mu_8),$ and $\mu_i = \frac{1}{2\pi {\rm i}} \log y_i$, $\tau = \frac{1}{2 \pi {\rm i}}\log \mathfrak{q}$.  
One can also express the elliptic genus of two strings as combinations of the $E_8$ theta functions, hence the E-strings enjoy $E_8$ Weyl symmetry.

Now we implement the defect Higgsing. In a similar fashion, we set
\begin{align}
\tilde{A}=q^{-M} \tilde{y}_3^{-1} = \tilde{y}_8,
\label{def6dEstr}
\end{align}
which is consistent with \eqref{defHiggs} with the duality map \eqref{map}. 
The resulting elliptic genera of the E-strings with the defect are given by
\begin{subequations}
\begin{align}
\tilde{Z}_{(1)}^{\rm defect}  = & \,\frac{1}{2} \frac{\eta^2}{\theta^2_1(q)} \sum_{I=1}^4 \left( \prod_{l\in \mathcal{I}} \frac{\theta_I (\tilde{y_l})}{\eta} \right) \left( \frac{\theta_I (\tilde{z})}{\theta_I (q^M \tilde{z})}\right),
\\
\tilde{Z}_{(2)}^{\rm defect} =\,
& \frac{1}{2} \frac{\eta^4}{\theta^2_1(q)\theta^2_1(q^2)} \sum_{I=1}^4 \left(  \prod_{l\in \mathcal{I}} \frac{\theta_I (\tilde{y_l}q^{1/2})\theta_I (\tilde{y_l}q^{-1/2})}{\eta^2} \right)  \left( \frac{\theta_I (\tilde{z}q^{1/2})\theta_I (\tilde{z}q^{-1/2})}{\theta_I (q^{M+1/2} \tilde{z})\theta_I (q^{M-1/2}\tilde{z})}\right)
\nonumber \\
&
\hspace{5mm}
+ \frac{1}{4} \frac{\eta^4}{\theta^4_1(q)} \sum_{(I,J,K)\in S} \frac{ \theta^2_I(1)}{\theta^2_I(q)}\prod_{l\in \mathcal{I}} \frac{\theta_J (\tilde{y_l})\theta_K (\tilde{y_l})}{\eta^2} \left( \frac{\theta_J (\tilde{z})\theta_K (\tilde{z})}{\theta_J (q^{M} \tilde{z})\theta_K (q^{M}\tilde{z})}\right),
\end{align}
\end{subequations}
where $\tilde{z} = \tilde{y}_3$ is the position of the defect insertion.  

We note here that unlike the usual Higgsing, $\tilde{Z}_{(1)}^{\rm Defect}$ and $\tilde{Z}_{(2)}^{\rm Defect}$ now are not invariant under E$_8$ Weyl symmetry, rather they are invariant under the SO(16) Weyl symmetry,
\begin{align}
\tilde{y}_i \leftrightarrow \tilde{y}_j,
\qquad\qquad \tilde{y}_i \leftrightarrow \tilde{y}^{-1}_j.
\end{align}
This implies that the E$_8$ global symmetry of the  elliptic genus for the E-string is broken SO(16) due to the presence of the defects.

To see the symmetry breaking more explicitly, let us consider the defect Higgsing of the elliptic genus expressed by the characters given in \cite{Kim:2015fxa}. A part of the contributions to the elliptic genus of one string $\tilde{Z}_{(1)}$ can be expressed as the character,
\begin{align}
\tilde{Z}_{(1)}=
&\frac{1}{(1-q)(1-q^{-1})} 
 \frac{\tilde{\mathfrak{q}}^{1/2}}{(1-\tilde{A}^2)(1-\tilde{A}^{-2})} \left(\tilde{Z}^{{\rm SO, \, SU}}_{(1)} + \tilde{Z}^{{\rm SU}}_{(1)} \right)  + \mathcal{O}(\tilde{\mathfrak{q}}^{3/2}),
 \label{6dSp1}
 \end{align}
where we decompose $\tilde{Z}_{(1)}$ into two parts: the contributions involving SO(20) characters and those not depending on SO(20) characters,
\begin{subequations}
 \begin{align}
 \tilde{Z}^{{\rm SO, \, SU}}_{(1)} &= 
-\chi_{\overline{{\bf 512}}}^{{\rm SO(20)}} \chi_{1/2}^{{\rm SU(2)}}(\tilde{A})
+2\chi_{{\bf 512}}^{{\rm SO(20)}} 
+2\chi_{{\bf 20}}^{{\rm SO(20)}} \chi_{3/2}^{{\rm SU(2)}}(\tilde{A})
-4\chi_{{\bf 20}}^{{\rm SO(20)}} \chi_{1/2}^{{\rm SU(2)}}(\tilde{A})
\nonumber \\
&\qquad\qquad
-\chi_{{\bf 190}}^{{\rm SO(20)}} \chi_{1}^{{\rm SU(2)}}(\tilde{A}) + 3\chi_{{\bf 190}}^{{\rm SO(20)}},
 \label{6dSp1A}
 \\
 \tilde{Z}^{{\rm SU}}_{(1)} &= 
6  \chi_{1/2}^{{\rm SU(2)}}(q) - 2  \chi_{1}^{{\rm SU(2)}}(\tilde{A})  \chi_{1/2}^{{\rm SU(2)}}(q) -5  \chi_{1}^{{\rm SU(2)}}(\tilde{A}) -3  \chi_{2}^{{\rm SU(2)}}(\tilde{A}),
 \label{6dSp1B}
\end{align}
\end{subequations}
where relevant the SO(20) characters are given by
\begin{subequations}
\begin{align}
&\chi_{{\bf 512}}^{{\rm SO(20)}} = \frac{1}{2} \left(\prod_{i=1}^{10}(\tilde{y}_i^{1/2}+\tilde{y}_i^{-1/2}) + \prod_{i=1}^{10}(\tilde{y}_i^{1/2} - \tilde{y}_i^{-1/2}) \right),
\\
&\chi_{\overline{{\bf 512}}}^{{\rm SO(20)}} = \frac{1}{2} \left(\prod_{i=1}^{10}(\tilde{y}_i^{1/2}+\tilde{y}_i^{-1/2}) - \prod_{i=1}^{10}(\tilde{y}_i^{1/2} - \tilde{y}_i^{-1/2}) \right),
\\
&\chi_{{\bf 190}}^{{\rm SO(20)}} = \sum_{1\leq i < j \leq 10} \left( \tilde{y}_i + \tilde{y}_i^{-1} \right) \left( \tilde{y}_j + \tilde{y}_j^{-1} \right) + 10,
\\
&\chi_{{\bf 20}}^{{\rm SO(20)}} = \sum_{i=1}^{10} \left( \tilde{y}_i + \tilde{y}_i^{-1} \right),
\end{align}
\end{subequations}
as well as the SU(2) character,
\begin{align}
\chi_{n}^{{\rm SU(2)}}(x) = \frac{x^{2n+1}-x^{-2n-1}}{x-x^{-1}}.
\end{align}

By the defect Higgsing \eqref{def6dEstr}, the part of the elliptic genus of one string \eqref{6dSp1A} reduces to
\begin{align}
\tilde{Z}^{{\rm SO\, SU}}_{(1)} &\to
\biggl( 
- \left(q^M \tilde{z} -q^{-M} {\tilde{z}}^{-1} \right) \left(q^{M/2}{\tilde{z}} - q^{-M/2}{\tilde{z}}^{-1} \right) \chi_{{\bf 128}}^{{\rm SO(16)}} 
\nonumber \\
&\qquad
- \left(q^M \tilde{z} -q^{-M} {\tilde{z}}^{-1} \right) \left(q^{M/2} - q^{-M/2} \right) \chi_{\overline{{\bf 128}}}^{{\rm SO(16)}} 
 -\left( q^{M}{\tilde{z}} - q^{-M}{\tilde{z}}^{-1} \right)^2  \chi_{{\bf 120}}^{{\rm SO(16)}} 
  \nonumber \\
 &\qquad
+\left( q^{M}{\tilde{z}} - q^{-M}{\tilde{z}}^{-1} \right)^2 \left(q^{M/2}{\tilde{z}} - q^{-M/2}{z'}^{-1} \right) \left(q^{M/2} - q^{-M/2} \right) \chi_{{\bf 16}}^{{\rm SO(16)}}
\biggr),
 \label{defSp1ch}
\end{align}
where we further introduce the SO(16) characters,
\begin{subequations}
\begin{align}
&\chi_{{\bf 128}}^{{\rm SO(16)}} = \frac{1}{2} \left(\prod_{i\in \mathcal{I}}(\tilde{y}_i^{1/2}+\tilde{y}_i^{-1/2}) + \prod_{i\in \mathcal{I}}(\tilde{y}_i^{1/2} - \tilde{y}_i^{-1/2}) \right),
\\
&\chi_{\overline{{\bf 128}}}^{{\rm SO(16)}} = \frac{1}{2} \left(\prod_{i\in \mathcal{I}} (\tilde{y}_i^{1/2}+\tilde{y}_i^{-1/2}) - \prod_{i\in \mathcal{I}}(\tilde{y}_i^{1/2} - \tilde{y}_i^{-1/2}) \right),
\\
&\chi_{{\bf 120}}^{{\rm SO(16)}} = \sum_{i,j\in \mathcal{I} ,i < j } \left( \tilde{y}_i +\tilde{y}_i^{-1} \right) \left( \tilde{y}_j + \tilde{y}_j^{-1} \right) ,
\\
&\chi_{{\bf 16}}^{{\rm SO(16)}} = \sum_{i\in \mathcal{I}} \left( \tilde{y}_i + \tilde{y}_i^{-1} \right).
\end{align}
\label{SO(16)ch}
\end{subequations}
When we set $M=0$ corresponding to the Higgsing, we can combine $\chi_{{\bf 128}}^{{\rm SO(16)}}$ and $\chi_{{\bf 120}}^{{\rm SO(16)}}$ into ${\rm E}_8$ character,
\begin{align}
-\left( \tilde{z} - {\tilde{z}}^{-1} \right)^2 \left(\chi_{{\bf 128}}^{{\rm SO(16)}}+\chi_{{\bf 120}}^{{\rm SO(16)}} \right)
=- \left( \tilde{z} - {\tilde{z}}^{-1} \right)^2 \chi_{{\bf 248}}^{{\rm E_8}} ,
\end{align}
which is nothing but the symmetry of E-string, Weyl symmetry of $E_8$.  In order to enhance the symmetry, the coefficients of $\chi_{{\bf 128}}^{{\rm SO(16)}}$ and $\chi_{{\bf 120}}^{{\rm SO(16)}}$  in \eqref{defSp1ch} have to be the same, whereas those of $\chi_{\overline{{\bf 128}}}^{{\rm SO(16)}}$ and $\chi_{{\bf 16}}^{{\rm SO(16)}}$ have to vanish, however, there is no such solution except for trivial cases, $M=0$. Therefore, we conclude that the symmetry of E-string is broken to ${\rm SO(16)}$ by introducing the defect.

\bigskip
\section{Global symmetry}\label{sec:globsymm}
In this section, we comment on global symmetry in the presence of codimension 2 defects. In computing the defect partition function, we used geometric transitions, where we construct ``unHiggsed'' 5-brane configuration and perform the defect Higgsing. As unHiggsed theory does not have any defect attached, the defects are introduced through defect Higgsing. Theories with defects that can be geometrically engineered hence captures some subalgebra of global symmetry of the unHiggsed theory. For instance, the partition function for the E-string theory with defects that we have computed in the previous section show manifest $\rm SO(16)$ global symmetry. We obtained it through the defect Higgsing of 6d Sp(1) gauge theory with 10 flavors which has ${\rm SO(20)}$ global symmetry. The defect of the E-string that we considered thus naturally has an ${\rm SO(16)}$ which can be understood as $\rm SO(20)$ global symmetry of the unHiggsed theory is broken to ${\rm SO(16)}$  along with the defect Higgsing. 

\begin{figure}
\centering
\includegraphics[width=13cm]{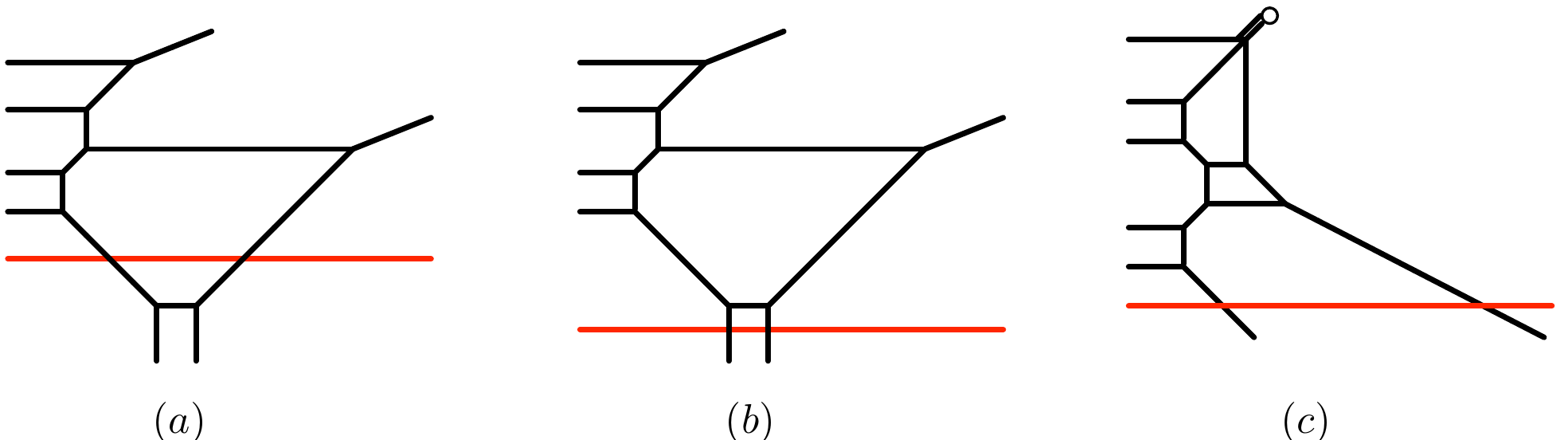}
\caption{Different unHiggsed 5-brane configurations for SU(2) gauge theory with 4 flavors and defects, associated with different  global symmetries: (a) SU(4)$\times$SU(2)$\times$SU(2), (b) SU(4)$\times$U(1)$\times$SU(2), and (c) SU(5)$\times$U(1). The red lines are D5-branes where the defect Higgsing is applied. A white dot is 7-brane.}
\label{fig:su2+4F}
\end{figure}

In M-theory perspective, $E_8$ global symmetry of E-string is the symmetry that arises as E-string probes M9 brane. Defects on E-string are expected not to disturb M9 brane, and hence one may expect to still see an $E_8$ symmetry for E-string theory with defects. Global symmetry of defect partition function, however, may depend on how we compute the defects. For instance, we can introduce different codimension 2 defects via geometric transitions for a given 5d theory by considering defect Higgsings on different unHiggsed 5-brane webs. These different defect Higgsings may capture different global symmetry, which could be the perturbative symmetry or more than the perturbative symmetry. As an instructive example, let us consider SU(2) theory with 4 flavors, whose perturbative symmetry is SO(8)$\times$U(1). In Figure \ref{fig:su2+4F}, three different unHiggsed 5-brane configurations are considered and each gives rise to different manifest global symmetries as depicted. 
The red lines in Figure \ref{fig:su2+4F} are D5-branes which we perform the defect Higgsing. The 5-brane configurations (a) and (b) are, in particular, instructive. Both unHiggsed 5-brane webs are the same as SU(3)$_0$ gauge theory with 6 flavors, their resulting global symmetries however are different. Global symmetry for (a) is SU(4)$\times$SU(2)$\times$SU(2), while that for (b) is SU(4)$\times$U(1)$\times$SU(2). Here SU(4) is the symmetry coming from the interchange of four flavor masses, and SU(2) arises from two parallel (2, 1) 5-branes in Figure~\ref{fig:su2+4F} (a) and (b), which is associated with instanton. Now the difference can be explained as follows. In (a), the defect Higgsing is applied to the internal D5-brane, which leaves two parallel (0,1) 5-branes on the bottom symmetric, accounting for the SU(2) part of its global symmetry. In (b), on the other hand, the defect Higgsings is applied to an outer D5-brane which makes two parallel (0,1) 5-branes on the bottom distinguishable, hence breaking SU(2) into U(1). The unHiggsed 5-brane configuration (c) is also noticeable. The resulting global symmetry is SU(5)$\times$U(1), which is obtained from an SU(4)$_{-\frac12}$ gauge theory with 7 flavors, with usual Higgsing in the upper part and also with a defect Higgsing in the lower part. From these three examples, it is clear that global symmetry of defect Higgsings comes from the global symmetry structure of unHiggsed 5-brane web, which could be simply perturbative symmetry or one of the maximal compact subgroup of the enhanced global symmetry, SO(10) in this particular example.  

As there could be many more unHiggsed 5-brane configurations that one can geometrically engineer, we expect different global symmetry depending on unHiggsed 5-brane that we used for the computations.


\section{Conclusion}\label{sec:conclusion}

In this paper, we computed the defect partition function of E-string theory on a circle, from 5-brane webs by applying the defect Higgsing, based on two 5-brane configurations: one with two O5-planes and the other without O5-planes. Though two 5-brane configurations look different, both parameter phases actually describe 5d SU(3)$_0$ gauge theory with 10 flavors, as the 5-brane web with two O5-planes is deformed to describe SU(3) gauge theory phases. We however referred to the one with O5-planes as Sp(2) gauge theory with 10 flavors just to distinguish natural 5-brane web, without orientifolds, for SU(3) gauge theory with 10 flavors. As  6d E-string theory on a circle is realized as 5d Sp(1) gauge theory with 8 flavors, we applied a suitable defect Higgsing on these 5-brane webs to obtain the surface defect partition function for 5d Sp(1) gauge theory with 8 flavors. The resulting partition function has the defect modulus which corresponds to the position on 5-brane where the defect is inserted. With the defect modulus, the partition function can be understood as the open topological string partition function. We compared our defect partition function with 6d elliptic genus result where the same defect Higgsing is implemented. We carried out our computation of the defect partition function up to 2-instantons and confirmed that the results agree up to that order.

For the 5-brane configurations that we considered in the paper, the insertion of the defect breaks global symmetry from $E_8$ to SO(16). This can be explicitly seen from the defect partition function obtained from the elliptic genus where the theta function at a given instanton order is invariant under the SO(16) Weyl transformation rather than the $E_8$ transformation. From the partition function of 5d Sp(1) gauge theory with 8 flavors, the SO(16) corresponds to the perturbative SO(16) flavor symmetry. The corresponding 5-brane web suggests that flavors are not affected by the defect and hence 8 flavor branes can be put closer to one of O5-planes so that they enjoy an SO(16) symmetry. In this way, one can see that regardless of the number of the defects, SO(16) global symmetry remains. Together with U(1) coming from the KK modes, the defect partition function hence has SO(16) $\times$ U(1)$_{KK}$ symmetry.

We note that there are different ways of realizing 5d brane system with defects, as discussed in section \ref{sec:globsymm}. Depending on how we implement the defect Higgsing, we may see different global symmetry structures. It is however that such apparent global symmetries are in fact a subgroup of the enhanced global symmetry, as one can readily restore the enhanced global symmetry when setting the number of defects to zero. Even for the case of the E-string theory with defects, global symmetry may restore to be $E_8$ if there is a nontrivial unHiggsed 5-brane configuration or nontrivial reparameterization involving the defect modulus such that it respects $E_8$ Weyl transformation. It is hence interesting to further study how the global symmetry structure is broken or hidden in the presence of defects. For instance, quantization of SW curve or associated integrable system \cite{Ruijsenaars_2015, Nazzal:2018brc, Haghighat:2018dwe, Chen:2020jla}. 

Some defects can be taken away by taking the defect insertion to infinity. In our setup, however, as we inserted the defect in between two NS5 branes, it is not possible to decouple the defects. Instead, one can set the number of the defects to zero, then one naturally recovers the E-string partition on a circle which restores $E_8$ global symmetry.

It is straightforward to generalize our computation to higher rank gauge theories. For instance, 6d $\mathcal{N}=1$ Sp($N$) gauge theory with $2N+8$ flavors with a tensor compactified on a circle can be realized as a 5-brane web with two O5-planes describing 5d $\mathcal{N}=1$ Sp($N+1$) gauge theory with $2N+8$ flavors. As we can apply the defect Higgsing to it, we expect to obtain 5d Sp($N$) gauge theory with $2N+6$ flavors and defects. Global symmetry for the theory with defects would be an SO($4N+12$) $\times$ U(1)$_{KK}$, as the corresponding 5-brane configuration suggests that there are O5-planes with $2N+6$ flavor D7-branes, and the instanton gives a U(1).

We can also consider the defect Higgsing for the system with fewer flavors. Starting from a 5d KK theory with defects, say 5d $\mathcal{N}=1$ Sp($N+1$) gauge theory with $2N+8$ flavors and defects, 
 one can decouple flavors one by one to get 5d $\mathcal{N}=1$ Sp($N+1$) gauge theory with $N_f<2N+8$ flavors and defects. For example, we worked out explicitly the partition function for 5d Sp(2) gauge theory with 9 and 8 flavors in Appendix \ref{decoupling}.


\bigskip

\acknowledgments
We thank Joonho Kim, Hee-Cheol Kim, and Kimyeong Lee for useful discussions. 
SSK thanks Interdisciplinary Center for Theoretical Study, University of Science and Technology of China for hospitality, KIAS and POSTECH for his visit where part of work is done, and also thanks APCTP for hosting the Focus program ``Strings, Branes and Gauge Theories.''  
YS thanks University of Electronic Science and Technology of China and South West Jiaotong University for hospitality for his visit.
YS is supported by a grant from the NSF of China with Grant No: 11675167 and 11947301.
FY is supported by the NSFC grant No. 11950410490, the Fundamental Research Funds for the Central Universities A0920502051904-48, by Startup research grant A1920502051907-2-046, in part by NSFC grant No. 11501470 and No. 11671328, and by Recruiting Foreign Experts Program No. T2018050 granted by SAFEA.
\bigskip

\appendix

\section{Conventions and Notations}\label{sec:appendix}
\subsection{Topological vertex}\label{sec:notation}
\begin{figure}[htb]
\centering
\includegraphics[width=4.5cm]{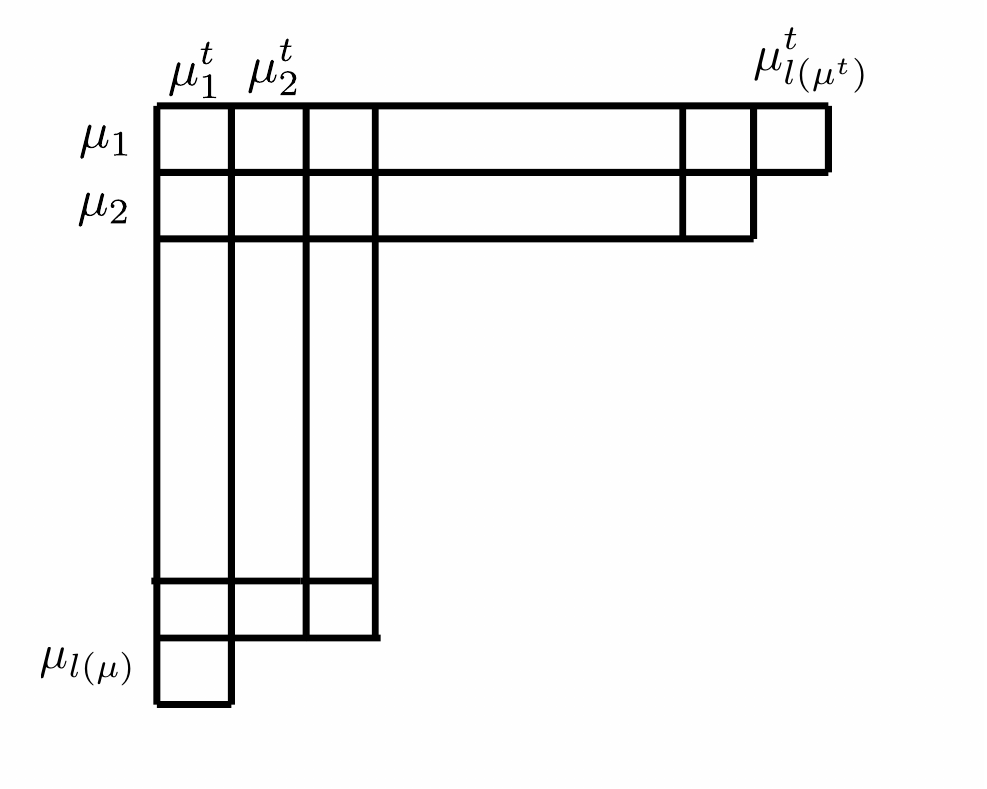}
\caption{The Young diagram and its parameters.}
\label{young}
\end{figure}

\begin{figure}[htb]
\centering
\includegraphics[width=5cm]{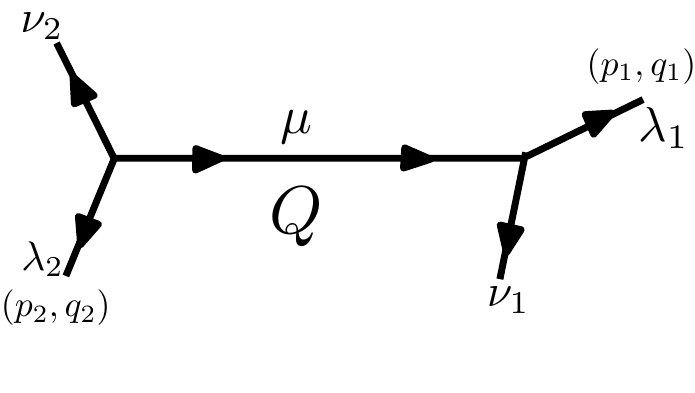}
\caption{The gluing rule of two vertices.}
\label{glue}
\end{figure}

Here we provide some definition and notation we use in this paper. We first define the Young diagram as in Figure \ref{young}.
The variables $\mu_i$, $\mu^t_j$, $l(\mu)$ denote the number of boxes in the $i$-th row, the number of boxes in the $j$-th column, and the number of the columns.  We also define the following quantities,
\begin{align}
|\mu| = \sum_i^{l(\mu)} \mu_i,\qquad ||\mu||^2=\sum_{i}^{l(\mu)} \mu_i^2,\qquad ||{\mu^t}||^2=\sum_{i}^{l(\mu)} {\mu^t}_i^2.
\end{align}
Based on the notation of the Young diagram, we define the topological vertex $C_{\lambda \mu \nu}$ \cite{Aganagic:2003db},
\begin{align}
C_{\lambda \mu \nu} &= q^{\frac{ ||\mu||^2 -||\mu^t||^2}{2}}s_{\nu^t}(q^{-\rho})
\sum_{\eta} s_{\lambda^{t}/\eta}(q^{-\nu-\rho})s_{\mu/\eta}(q^{-\rho-\nu^{t}}),
\end{align}
where $\lambda, \mu, \nu$ and $\eta$ are the Young diagrams, and $q$ relates to the string coupling, $q={\rm e}^{{\rm i}g_s}$. The function $s_{\mu}(x)$ is {\it Schur function} defined by,
\begin{align}
s_\mu (x_1,...,x_N) = \frac{\det x_j^{\mu_i+N-i}}{\det x_j^{N-i}},
\end{align}
and $s_{\mu/\eta}(x)$ is {\it skew Schur function} defined by the summation of the Schur function with the weight $N^{\mu}_{\nu\eta}$ called as {\it Littlewood--Richardson coefficients},
\begin{align}
s_{\mu/\eta}(x) = \sum_{\nu} N^{\mu}_{\nu\eta} s_{\nu}(x).
\end{align}
\begin{figure}[htb]
\centering
\includegraphics[width=3cm]{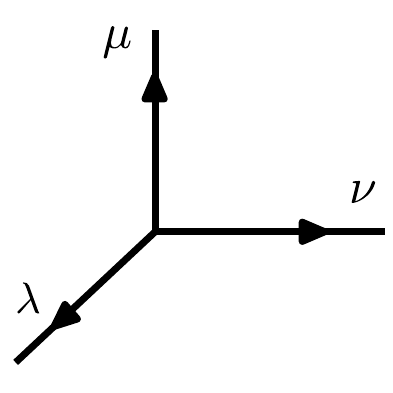}
\caption{The pictorial description of the topological vertex.}
\label{vertex}
\end{figure}
We also define the {\it framing factor} $f_\mu$,
\begin{align}
f_\mu =(-1)^{|\mu|} q^{-\frac{||\mu||^2-||\mu^t||^2}{2}}.
\label{framingfactor}
\end{align}
From these ingredients, the gluing rule for two vertices is given by (see also Figure \ref{glue}),
\begin{align}
\sum_{\mu} C_{\lambda_1 \nu_1 \mu}  C_{\lambda_2 \nu_2 \mu^t} (-Q)^{|\mu|} f_{\mu}^{n},~n=(p_1,q_1)\wedge (p_2,q_2).
\label{glue1}
\end{align}
By gluing some vertices, we can calculate a partition function of topological strings on non-compact toric Calabi--Yau manifolds. One can find more detailed computations and some examples in  \cite{Aganagic:2003db}.


\subsection{Theta function}\label{sec:thetaFunction}
To provide the elliptic genus of E-strings, we provide the definition of Jacobi's theta function,
\begin{subequations}
\label{eq:JacobiTheta}
\begin{align}
&\theta_1(p;y) := - {\rm i} p^{1/8} y^{1/2} \prod_{n=1}^\infty (1-p^n) (1- p^n y)(1-p^{n-1}y^{-1} ) ,
\\
&\theta_2 (p;y) =  p^{1/8} y^{1/2} \prod_{n=1}^\infty (1-p^n) (1+p^n y)(1+p^{n-1}y^{-1}),
\\
&\theta_3 (p;y) = \prod_{n=1}^\infty (1-p^n) (1+p^{n-1/2} y)(1+p^{n-1/2}y^{-1}),
\\
&\theta_4 (p;y) = \prod_{n=1}^\infty (1-p^n) (1-p^{n-1/2} y)(1-p^{n-1/2}y^{-1}),
\end{align}
\end{subequations}
They satisfy the following reflection formula,
\begin{align}
\theta_I (p;y^{-1}) = (-1)^{\delta_{I,1}} \theta_I(p;y),
\end{align}
where  $\delta_{I,j}$ is Kronecker delta, which is used to show that the elliptic genus of 6d Sp$(1)$ gauge theory reduces to the one of E-strings after the Higgsing. In the section \ref{elliptic}, we use short notation,
\begin{align}
\theta_I (\tilde{\mathfrak{q}};y) = \theta_I(y),
\end{align}
where $\tilde{\mathfrak{q}}$ is instanton fugacity in 6d theory relating to the one in 5d theory $\mathfrak{q}$, $\tilde{\mathfrak{q}}=\mathfrak{q}^2$, as we denoted in \eqref{map}.

\subsection{Plethystic exponential}
To express the physical part of the partition functions, it is useful to express them as plethystic exponential defined by,
\begin{align}
{\rm PE}\left[ f({\bm x}) \right] = {\rm exp} \left[\sum_{n=1}^\infty \frac{1}{n} f({\bm x}^n) \right],
\end{align}
where ${\bm x} = \{x_1, x_2,... \}$ and ${\bm x}^n = \{x_1^n, x_2^n,... \}$
For instance, an infinite product $\prod_{i,j=1}^\infty (1-Q q^{i+j-1})$ can be expressed as
\begin{align}
\prod_{i,j=1}^\infty (1-Q q^{i+j-1}) = {\rm exp}\left[\sum_{n=1}^\infty \frac{1}{n} \frac{q^n Q^n}{(1-q^{n})^2} \right] = {\rm PE} \left[\frac{q}{(1-q)^2}Q \right].
\end{align}
From this expression, the analytic continuation formula which can be understood as flop transition is given by
\begin{align}
{\rm PE}\left[ \frac{q}{(1-q)^2} Q \right] \to {\rm PE} \left[\frac{q}{(1-q)^2}  Q^{-1} \right].
\label{flop}
\end{align}
This formula is sometimes used to express the partition function as characters.


\section{K\"ahler parameters of the Tao diagram}
\label{app:Kahler}
\begin{figure}[htb]
\centering
\includegraphics[width=6cm]{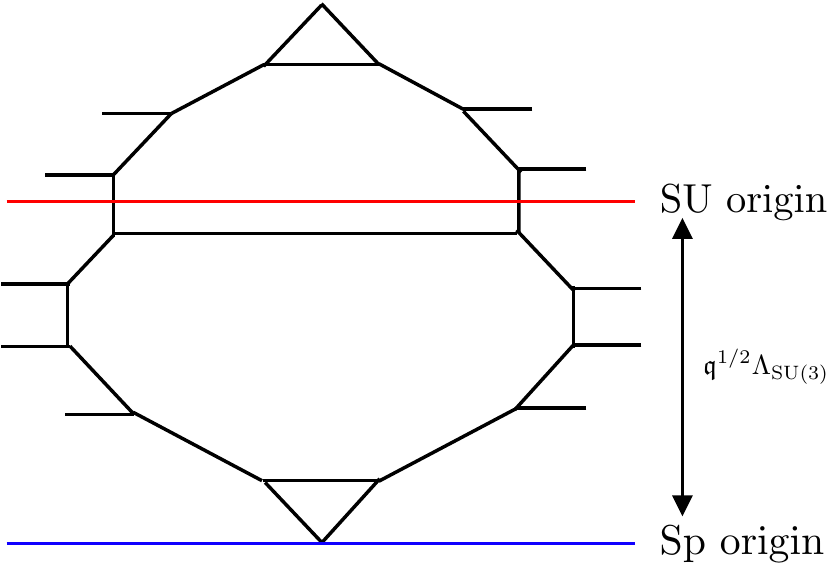}
\caption{The difference of the SU origin and the Sp origin. In this figure we omit O5-plane.}
\label{SpandSU}
\end{figure}

Here, we summarize the expression for the K\"ahler parameters of the Tao diagram depicted in Figure \ref{fig:tao} in terms of the parameters of the 5d $SU(3)$ gauge theory with $10$ flavors.
The derivation detail is given in the appendix of \cite{Hayashi:2016abm}.
The K\"ahler parameters associated with the middle strip are given by
\begin{align}
&Q_{1} = \frac{M_1}{A_1},\qquad
\tilde{Q}_{1} = \frac{A_1}{M_2},\qquad
Q_{2} = \frac{M_2}{A_2},\qquad
\tilde{Q}_{2} = \frac{A_2}{M_3},
\cr
&Q_{3} = \frac{M_3}{A_3},\qquad
\tilde{Q}_{3} = \frac{A_3}{M_5},\qquad
Q_{4} 
=  \mathfrak{q} \sqrt{ \frac{M_4 M_5}{M_1  M_2 M_3 M_6 M_7 M_8 M_9 M_{10}} },
\cr
&Q_{5} = \frac{A_3}{M_6},\qquad
\tilde{Q}_{5} = \frac{M_7}{A_3},\qquad
Q_{6} = \frac{A_2}{M_7},\qquad
\tilde{Q}_{6} = \frac{M_8}{A_2},
\cr
&Q_{7} = \frac{A_1}{M_8},\qquad
\tilde{Q}_{7} = \frac{M_{10}}{A_1},\qquad
Q_{8} = \mathfrak{q} \sqrt{ \frac{M_6 M_7 M_8 M_1 M_2 M_3 M_4 M_{5}}{M_9 M_{10} } },
\cr
&Q_{b_1} =   \mathfrak{q} A_1\sqrt{
\frac{
M_2 M_3 M_4 M_5 M_6 M_7 M_8}{M_1 M_9  M_{10}} },
\qquad
Q_{b_2} =  \mathfrak{q}  \frac{A_1}{A_3}\sqrt{
\frac{
 M_3 M_4 M_5 M_6 M_7}{
 M_1 M_2 M_8 M_9 M_{10} } },
\cr
&Q_{b_3} =   \frac{\mathfrak{q}}{A_3}\sqrt{
 \frac{M_4 M_5 M_6}{
  M_1M_2M_3M_7M_8M_9M_{10} } }\,
\end{align}
whereas those for the spiral strip 1 and 2 are given by
\footnotesize
\begin{align}
&
Q^{(1)}_1 = \frac{1}{M_2 M_3 M_4},
\quad 
Q^{(1)}_2 = \frac{M_1}{M_5},
\quad 
Q^{(1)}_3 =\mathfrak{q} \sqrt{\frac{M_1 M_5}{M_2 M_3 M_4 M_6 M_7 M_8 M_9 M_{10} }},
\quad
Q^{(1)}_4 =\mathfrak{q} \sqrt{\frac{M_1 M_6 M_7 M_8 M_9}{M_2 M_3 M_4 M_5 M_{10} }},
\cr
&
Q^{(1)}_5 =\mathfrak{q} \sqrt{\frac{M_1 M_{10}}{M_2 M_3 M_4 M_5 M_6 M_7 M_8 M_9}},
\quad
Q^{(1)}_6 = \mathfrak{q}^2 \frac{M_1}{M_{10}},
\quad 
Q^{(1)}_7 = \mathfrak{q}^2 \frac{1}{M_2 M_3 M_4},
\quad 
Q^{(1)}_8= \mathfrak{q}^2 \frac{M_1}{M_5},
\cr
&
Q^{(2)}_1 = \frac{1}{A_1 M_3 M_4},
\quad 
Q^{(2)}_2 = \frac{M_2}{M_5},
\quad 
Q^{(2)}_3 = \mathfrak{q} \sqrt{ \frac{M_2 M_5}{ M_1 M_3 M_4 M_6 M_7 M_8 M_9 M_{10} }},
\quad
Q^{(2)}_4 = \mathfrak{q} \sqrt{\frac{M_2 M_6 M_7 M_8 M_9}{M_1 M_3 M_4 M_5 M_{10} } },
\cr
&
Q^{(2)}_5 = \mathfrak{q} \sqrt{ \frac{M_2 M_{10} }{M_1 M_3 M_4 M_5 M_6 M_7 M_8 M_9} },
\quad
Q^{(2)}_6 = \mathfrak{q}^2 \frac{M_2}{M_{10}},
\quad 
Q^{(2)}_7 = \mathfrak{q}^2 \frac{1}{M_1 M_3 M_4},
\quad 
Q^{(2)}_8= \mathfrak{q}^2 \frac{M_2}{M_5},
\cr
&
Q^{(3)}_1 = \frac{A_3}{M_4},
\quad 
Q^{(3)}_2 = \frac{M_3}{M_5},
\quad 
Q^{(3)}_3 = \mathfrak{q} \sqrt{\frac{M_3 M_5}{M_1 M_2 M_4 M_6 M_7 M_8 M_9 M_{10} }},
\quad
Q^{(3)}_4 = \mathfrak{q} \sqrt{\frac{M_3 M_6 M_7 M_8 M_9}{M_1 M_2 M_4 M_5 M_{10} }},
\cr
&
Q^{(3)}_5 = \mathfrak{q} \sqrt{ \frac{M_3 M_{10} }{M_1 M_2 M_4 M_5 M_6 M_7 M_8 M_9} },
\quad
Q^{(3)}_6 = \mathfrak{q}^2 \frac{M_3}{M_{10}},
\quad 
Q^{(3)}_7 = \mathfrak{q}^2 \frac{1}{M_1 M_2 M_4},
\quad 
Q^{(3)}_8= \mathfrak{q}^2 \frac{M_3}{M_5},
\cr
&
\qquad  \qquad  \qquad  \qquad \qquad  \qquad  \qquad  
Q^{(4)}_3 = \mathfrak{q} \sqrt{ \frac{M_4 M_5}{M_1 M_2 M_3 M_6 M_7 M_8 M_9 M_{10} }},
\quad
Q^{(4)}_4 = \mathfrak{q} \sqrt{ \frac{M_4 M_6 M_7 M_8 M_9}{M_1  M_2 M_3 M_5 M_{10}} },
\cr
&
Q^{(4)}_5 =\mathfrak{q} \sqrt{ \frac{M_4 M_{10} }{M_1 M_2 M_3 M_5 M_6 M_7 M_8 M_9} },
\quad
Q^{(4)}_6 = \mathfrak{q}^2 \frac{M_4}{M_{10}},
\quad 
Q^{(4)}_7 = \mathfrak{q}^2 \frac{1}{M_1 M_2 M_3},
\quad 
Q^{(4)}_8= \mathfrak{q}^2 \frac{M_4}{M_5},
\cr
&
Q^{(5)}_1 = M_7 M_8 M_9,
\quad 
Q^{(5)}_2 = \frac{M_{10}}{M_6},
\quad 
Q^{(5)}_3 = \mathfrak{q} \sqrt{ \frac{M_1 M_2 M_3 M_4 M_5 M_7 M_8 M_9}{M_6 M_{10}} }, 
\quad
Q^{(5)}_4 = \mathfrak{q} \sqrt{ \frac{ M_5 M_7 M_8 M_9 M_{10}}{M_1 M_2 M_3 M_4 M_6} },
\cr
&
Q^{(5)}_5 = \mathfrak{q} \sqrt{ \frac{M_1 M_2 M_3 M_4 M_7 M_8 M_9 M_{10} }{M_5 M_6}},
\quad
Q^{(5)}_6 = \mathfrak{q}^2 \frac{M_5}{M_6},
\quad 
Q^{(5)}_7 = \mathfrak{q}^2 M_7 M_8 M_9,
\quad 
Q^{(5)}_8= \mathfrak{q}^2 \frac{M_{10}}{M_6},
\cr
&
Q^{(6)}_1 = A_3 M_8 M_9,
\quad 
Q^{(6)}_2 = \frac{M_{10}}{M_7},
\quad 
Q^{(6)}_3 = \mathfrak{q} \sqrt{ \frac{M_1 M_2 M_3 M_4 M_5 M_6 M_8 M_9}{M_7 M_{10}} },
\quad
Q^{(6)}_4 = \mathfrak{q} \sqrt{ \frac{M_5 M_6 M_8 M_9 M_{10} }{M_1 M_2 M_3 M_4 M_7} },
\cr
&
Q^{(6)}_5 = \mathfrak{q} \sqrt{ \frac{M_1 M_2 M_3 M_4 M_6 M_8 M_9 M_{10}}{M_5 M_7} },
\quad
Q^{(6)}_6 = \mathfrak{q}^2 \frac{M_5}{M_7},
\quad 
Q^{(6)}_7 = \mathfrak{q}^2 M_6 M_8 M_9,
\quad 
Q^{(6)}_8= \mathfrak{q}^2 \frac{M_{10}}{M_7},
\cr
&
Q^{(7)}_1 = \frac{M_9}{A_1},
\quad 
Q^{(7)}_2 = \frac{M_{10}}{M_8},
\quad 
Q^{(7)}_3 = \mathfrak{q} \sqrt{ \frac{M_1 M_2 M_3 M_4 M_5 M_6 M_7 M_9}{M_8 M_{10}} },
\quad
Q^{(7)}_4 = \mathfrak{q} \sqrt{ \frac{M_5 M_6 M_7 M_9 M_{10}}{M_1 M_2 M_3 M_4 M_8} },
\cr
&
Q^{(7)}_5 =\mathfrak{q} \sqrt{ \frac{M_1 M_2 M_3 M_4 M_6 M_7 M_9 M_{10} }{M_5 M_8} } ,
\quad
Q^{(7)}_6 = \mathfrak{q}^2 \frac{M_5}{M_8},
\quad 
Q^{(7)}_7 = \mathfrak{q}^2 M_6 M_7 M_9,
\quad 
Q^{(7)}_8= \mathfrak{q}^2 \frac{M_{10}}{M_8},
\cr
&
\qquad  \qquad  \qquad  \qquad \qquad  \qquad  \qquad  
Q^{(8)}_3 = \mathfrak{q} \sqrt{ \frac{M_1 M_2 M_3 M_4 M_5 M_6 M_7 M_8}{M_9 M_{10} } },
\quad
Q^{(8)}_4 = \mathfrak{q} \sqrt{ \frac{M_5 M_6 M_7 M_8 M_{10} }{M_1 M_2 M_3 M_4 M_9} },
\cr
&
Q^{(8)}_5 =\mathfrak{q} \sqrt{ \frac{M_1 M_2 M_3 M_4 M_6 M_7 M_8  M_{10}}{M_5 M_9} },
\quad
Q^{(8)}_6 = \mathfrak{q}^2 \frac{M_5}{M_9},
\quad 
Q^{(8)}_7 = \mathfrak{q}^2 M_6 M_7 M_8,
\quad 
Q^{(8)}_8= \mathfrak{q}^2 \frac{M_{10}}{M_9}.
\end{align}
\normalsize
Here $M_i~(i=1,...,10)$ denote the position of horizontal lines measured from ``SU(3)" origin (see Figure \ref{SpandSU}).
To convert e.g. from Sp origin to SU origin, we have to multiply $\mathfrak{q}^{1/2}\Lambda_{\text{SU(3)}}$ for all variables $x_I$ and $M_i$.
We note that the other K\"ahler parameters $Q^{(i)}_{j} $ with large $j$ is obtained by
\begin{align}
Q^{(i)}_{j+6} = \mathfrak{q}^2 Q^{(i)}_{j} \qquad ({\rm for} \,\, i =1,2,\cdots, 8, \quad  j \ge 3).
\end{align}
These are all the K\"ahler parameters which are necessary to compute the $2$-instanton contribution.


\section{\texorpdfstring{$(p,q)$ 5-brane web with O5-planes and Topological vertex}{p,q}}

\subsection{Topological vertex formalism with O5-planes}\label{TVFO5}

Here we summarize the computation rule of the partition function of the Sp gauge theories realized by the web diagrams with O5-planes discussed in \cite{Kim:2017jqn}. We decompose the web diagram into the left side and right side about the red line. Then, we flip the sign of the first elements of the charge vectors, $(p,q)\to (-p,q)$, whose graphical meaning is to fold one side along the O5-plane. Here we fold the right side.
\begin{figure}[htb]
\centering
\includegraphics[width=7.5cm]{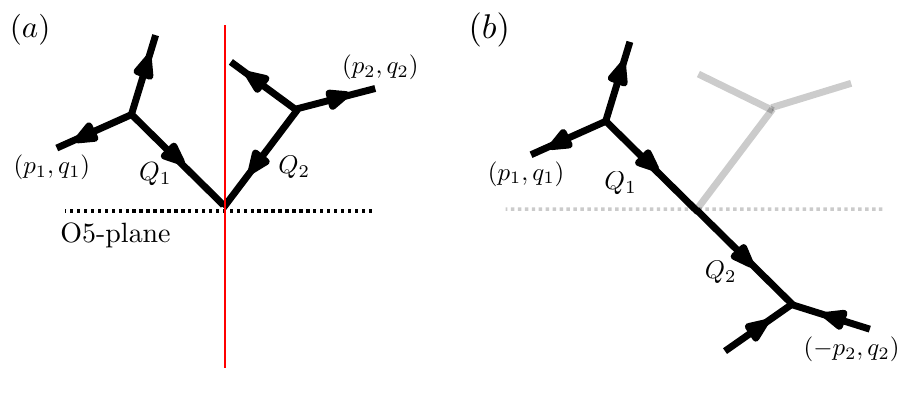}
\caption{The rule how to interpret the web diagram with the O5-plane depicted in Figure (a). With respect to the O5-plane, we flip the sign of the first elements of the charge vectors.}
\label{rule}
\end{figure}
The glueing rule of the vertices across the O5-plane is given as follows, 
\begin{align}
\sum_{\mu}C_{\lambda_1 \mu \nu_1} C_{\lambda_2 \mu^t \nu_2} (+Q_1 Q_2)^{|\mu|} f_\mu^{\mathfrak{n}},~\mathfrak{n} = (p_2,-q_2) \wedge (p_1, q_1)+1.
\end{align}

Now we are ready to write down the partition function of the Sp$(2)$ gauge theory with 10 flavors whose web diagram is given in Figure \ref{webO5},
\begin{align}
Z^{{\rm Sp(2)+10F}}
=
&\sum_{\{ \nu_i \},\{\mu_i \}}
\left(\frac{{\mathfrak{q}'}^2}{{x'}_1^2} \right)^{|\mu_1|} \left(\frac{\prod_{i=3}^5 y'_i y'_{i+5}}{{x'}_2^2 {x'}_3^2} \right)^{|\mu_2|} \left({x'}_3^2 \right)^{|\mu_3|} f^{-3}_{\mu_1} f^3_{\mu_3}
\nonumber \\
&\times(-{x'}_1 {y'}_1^{-1})^{|\nu_1|} (-y'_1 {y'}_2^{-1})^{|\nu_2|} (-y'_2 {x'}_2^{-1})^{|\nu_3|} (-x'_2 {y'}_3^{-1})^{|\nu_4|} (-y'_3 {y'}_4^{-1})^{|\nu_5|} (-y'_4 {y'}_5^{-1})^{|\nu_6|} 
\nonumber \\
&\times(-y'_5 {x'}_3^{-1})^{|\nu_7|}(-{x'}_3^2)^{|\nu_8|} (-y'_6 {x'}_3^{-1})^{|\nu_9|} (-y'_7 {y'}_6^{-1})^{|\nu_{10}|} (-y'_8 {y'}_7^{-1})^{|\nu_{11}|} \times(-x'_2 {y'}_8^{-1})^{|\nu_{12}|} 
\nonumber \\
&\times(-y'_9 {x'}_2^{-1})^{|\nu_{13}|} (-y'_{10} {y'}_9^{-1})^{|\nu_{14}|} (-{x'}_{1} {y'}_{10}^{-1})^{|\nu_{15}|} (-{\mathfrak{q}'}^2 {x'}_{1}^{-2})^{|\nu_{16}|}
\nonumber \\
&\times f_{\nu_2}^{-1} f_{\nu_5}^{-1} f_{\nu_6}^{-1} f_{\nu_{10}}^{-1}  f_{\nu_{11}}^{-1}  f_{\nu_{14}}^{-1} 
\nonumber \\
&\times C_{\nu_1 \nu_{16}^t \mu_1} C_{\nu_1^t \nu_2^t \emptyset} C_{\nu_2 \nu^t_3 \emptyset} C_{\nu_4^t \nu_3 \mu_2} C_{\nu_4 \nu_5^t \emptyset} C_{\nu_5 \nu_6^t \emptyset} C_{\nu_6 \nu_7^t \emptyset} C_{\nu_8 \nu_7 \mu_3} 
\nonumber \\
&\times C_{\nu_8^t \nu_9^t \mu_3^t} C_{\nu_{10} \nu_9 \emptyset} C_{\nu_{11} \nu_{10}^t \emptyset} C_{\nu_{12} \nu_{11}^t \emptyset} C_{\nu_{12}^t \nu_{13} \mu_2^t} C_{\nu_{14} \nu_{13}^t \emptyset} C_{\nu_{15}^t \nu_{14}^t \emptyset} C_{\nu_{15} \nu_{16} \mu_1^t}.
\end{align}
The efficient way to calculate the partition function is to use the operator formalism explained in \cite{Kimura:2018kaf}. After some computations, one finds obtain \eqref{Sp210F}. To reach the result, we use the analytic continuation formula,
\begin{align}
\Theta_{\mu\nu}(Q) \rightarrow Q^{|\mu|+|\nu|}f_\mu f_\nu \Theta_{\mu^t \nu^t}(Q^{-1}).
\end{align}

\subsection{The partition function of 
\texorpdfstring{SU$(N+1)_0$ + $(2N+6){\bf F}$}{SU(N+1)0 + (2N+6)F}
}
In section \ref{sec;equivO5}, we have calculated the partition function of  5d $\mathcal{N}=1$ SU$(3)$ gauge theory with 10 flavors.
Then, after the usual Higgsing, we have found the partition function of the SU$(2)$ gauge theory with 8 flavors in section \ref{sec:HiggsingVSDefHiggsing}.
Based on these results, we propose the partition function of SU$(N+1)_0$ gauge theory with $2N+6$ flavors without any computations,
\begin{align}
&Z^{{\rm SU(}N{\rm +1)+(2}N{\rm +6)F}}
=
Z^{2N+6}_{{\rm mass}}\sum_{\mu_{1,...,N+1}} 
\prod_{I=1}^{N+1}
\left( \mathfrak{q} \Lambda^{-2}_{{\rm SU}(N+1)} A_I^{-3-N} \right)^{|\mu_I|}
f_{\mu_I}^{-3-N}
\nonumber \\
&\hspace{35mm}\times \prod_{I=1}^{N+1} \left(\prod_{i=1}^{2N+6} \frac{\Theta_{\mu_I {\Emptyset}}(A_I M_i^{-1})}{\Theta_{\mu_I {\Emptyset}}(\mathfrak{q} A_I M_i \Lambda^2_{{\rm SU}(N+1)})} \prod_{J=1}^{N+1} \frac{\Theta_{\mu_I \mu_J}(\mathfrak{q} A_I A_J \Lambda^2_{{\rm SU}(N+1)} )}{\Theta_{\mu_I \mu_J^t}(A_I A_J^{-1})} \right),
\\
&Z^{2N+6}_{{\rm mass}}=\prod_{i,j=1}^\infty \frac{1}{(q^{i+j-1}\mathfrak{q}:\mathfrak{q})_\infty^{N+3}} \times \frac{\prod_{1\leq i, j\leq N+3} \Theta_{{\Emptyset}{\Emptyset}}(\mathfrak{q} M_i M_{j+N+3} \Lambda^2_{{\rm SU}(N+1)})}{\prod_{1\leq i<j\leq N+3} \Theta_{{\Emptyset}{\Emptyset}}(M_i M^{-1}_j)\Theta_{{\Emptyset}{\Emptyset}}(M_{i+N+3} M^{-1}_{j+N+3})},
\end{align}
where $\Lambda_{{\rm SU}(N+1)} = \prod_{i=1}^{2N+6}M_i^{-1/4}$ and the Coulomb branch moduli satisfy $\prod_{I=1}^{N+1}A_{I,{\rm SU}(N+1)}=1$.


\section{Flavor decoupling limit}\label{decoupling} 

To consider what global symmetry is preserved in the presence of the defect in the 5d gauge theories with $N_f < 10$ flavors, we shall take the decoupling limit and consider the defect Higgsing. The gauge theory with 9 flavors can be obtained by the following decoupling limit,
\begin{align}
&y'_{10}\to0, \qquad
\mathfrak{q}' \to 0, \qquad
\frac{{\mathfrak{q}}'^2}{y'_{10}}={\mathfrak{q}'}^2_{9}~ \text{(fixed)},
\end{align}
\begin{figure}[htb]
\centering
\includegraphics[width=12cm]{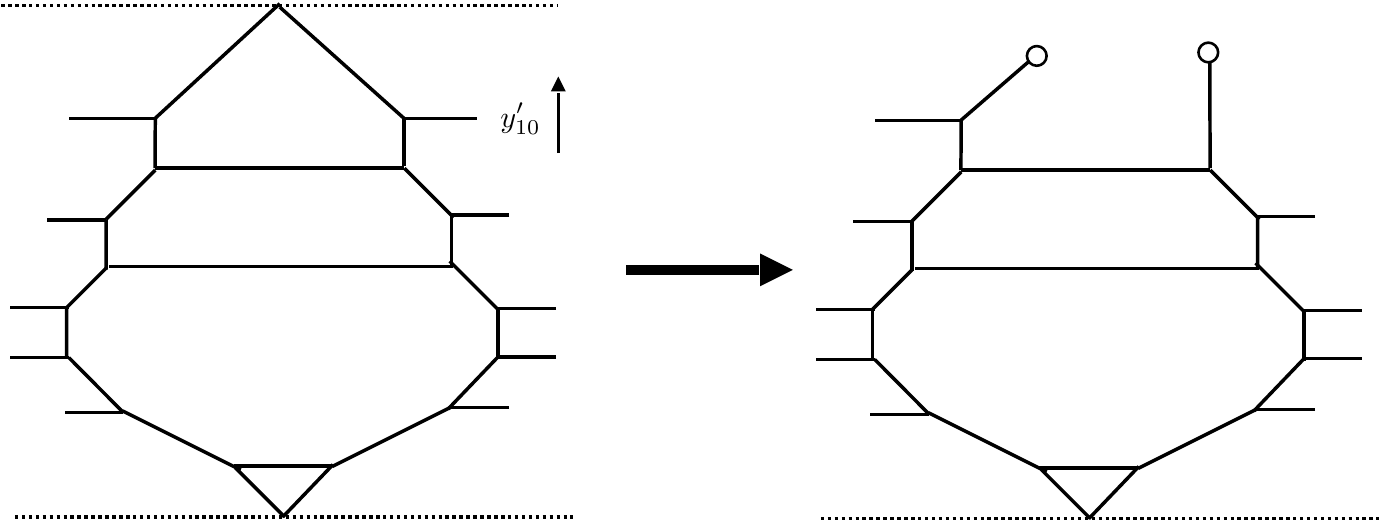}
\caption{A pictorial explanation of the decoupling limit from 10 flavors to 9 flavors.}
\label{webO5_decouple}
\end{figure}
\noindent
where $\mathfrak{q}'_{9}$ is the instanton factors of 5d Sp(2) gauge theory with 9 flavors. 
Diagrammatically, the decoupling limit is given in Figure \ref{webO5_decouple}.
The left side of this web diagram and the one in Figure \ref{webO5} are equivalent under the flop transition \cite{Iqbal:2004ne, Konishi:2006ev, Taki:2008hb}.
Then, the decoupling limit is done by removing one of the external legs from the web diagram, and the resulting web diagram is the right side of Figure \ref{webO5_decouple}, where the white circles denote the D7-branes. Since we remove one of the O5-plane, this limit also implies the decompactification limit.

From the elliptic genus of 6d Sp(1) gauge theory with 10 flavors \eqref{6dSp1} with the duality map \eqref{SpSp10F}, we have the partition function of 5d Sp(2) gauge theory with 9 flavors,
\begin{align}
{Z'}^{\rm{Sp(2)+9F}} =Z'_{(0),\text{9F}} + Z'_{(1),\text{9F}} {A'}_1 + \mathcal{O}({A'}_1^2) 
\end{align}
with
\begin{subequations} \label{Sp2_9F_0}
\begin{align}
&Z'_{(0),\text{9F}} = \frac{1}{2}  \frac{q}{(1-q)^2} \left( 2\left( {A'}_2^2 + {A'}_2^{-2} \right) -\left( {A'}_2 + {A'}_2^{-1} \right) \left( \sum_{i=1}^{9} \left( y'_i + {y'}_i^{-1} \right) +(\mathfrak{q}'+{\mathfrak{q}'}^{-1}) \right)\right),
\\
&Z'_{(1),\text{9F}}=
\frac{1}{(1-q)(1-q^{-1})} 
 \frac{{\mathfrak{q}'}_9^{2}}{(1-{A'}_2^2)(1-{A'}_2^{-2})} \left({Z'}^{{\rm SO \, SU}}_{(1),\text{9F}} + {Z'}^{{\rm SU}}_{(1),\text{9F}} \right),
\end{align}
\end{subequations}
where
\begin{subequations}
\begin{align}
&{Z'}^{{\rm SO, \, SU}}_{(1),\text{9F}} = \tilde{Z}^{{\rm SO, \, SU}}_{(1)}|_{\tilde{y}_{1,...,9}\to y'_{1,...,9},~ \tilde{y}_{10}\to \mathfrak{q}'_9,~\tilde{A}\to A'_2},
\\
&{Z'}^{{\rm SU} }_{(1),\text{9F}} = \tilde{Z}^{{\rm SU} }_{(1)}|_{\tilde{y}_{1,...,9}\to y'_{1,...,9},~ \tilde{y}_{10}\to \mathfrak{q}'_9,~\tilde{A}\to A'_2}
\end{align}
\end{subequations}
Note that the elliptic genus for 1-string \eqref{6dSp1} is given as a series of $\mathfrak{q}'$, however, \eqref{Sp2_9F_0} is exact results in $\mathfrak{q}'$.

Based on \eqref{SpSU10F}, we find the duality map between Sp(2) and SU(3) gauge theories with 9 flavors is
\begin{equation}
\begin{aligned}
&y'_i = \mathfrak{q}^{1/2} \left(\prod_{j=1}^{9}M_j^{-1/4} \right) M_i~(i=1,..,9),
\\
&A'_i = \mathfrak{q}^{1/2} \left(\prod_{j=1}^{9}M_j^{-1/4} \right) A_i (i=1,2),\quad
\mathfrak{q}'=\mathfrak{q}^{3/4}  \left(\prod_{j=1}^{9}M_j^{1/8} \right) .
\label{SpSU9F}
\end{aligned}
\end{equation}
We check the agreement of the partition functions between the Sp(2) and SU(3) gauge theories with 9 flavors under the duality map up to second order of $A_1$ and $\mathfrak{q}$. We also check that, by expanding the partition functions of Sp(2) and SU(3) gauge theory as a series of $A_1$ and $\mathfrak{q}$, and using the analytic continuation formula for \eqref{Sp2_9F_0}, these partition functions have only positive power of $A_1$ and $\mathfrak{q}$.

By the defect Higgsing given by,
\begin{align}
A'_2=q^M y'_3 = {y'}_8,
\label{def5dSp2}
\end{align}
which is derived from the Higgsing in 6d Sp(1) theory with 10 flavors  \eqref{def6dEstr} and the duality map \eqref{SpSp10F}, we obtain the Sp(1) gauge theory with 7 flavors in the presence of the defects, and ${Z'}^{{\rm SO, \, SU}}_{(1),9F}$ is given by
\begin{align}
{Z'}^{{\rm SO, \, SU}}_{(1),\text{9F}}\to
&
\biggl( 
- \left(q^M z' -q^{-M} {z'}^{-1} \right) \left(q^{M/2}{z'} - q^{-M/2}{z'}^{-1} \right) {\chi'}_{{\bf 128}}^{{\rm SO(16)}} 
\nonumber \\
&
- \left(q^M z' -q^{-M} {z'}^{-1} \right) \left(q^{M/2} - q^{-M/2} \right) {\chi'}_{\overline{{\bf 128}}}^{{\rm SO(16)}} 
 -\left( q^{M}{z'} - q^{-M}{z'}^{-1} \right)^2  {\chi'}_{{\bf 120}}^{{\rm SO(16)}} 
  \nonumber \\
 &
+\left( q^{M}{z'} - q^{-M}{z'}^{-1} \right)^2 \left(q^{M/2}{z'} - q^{-M/2}{z'}^{-1} \right) \left(q^{M/2} - q^{-M/2} \right) {\chi'}_{{\bf 16}}^{{\rm SO(16)}}
\biggr),
\end{align}
where $z' = y'_3$ is the brane moduli of 5d Sp(1) gauge theory. Here, prime of $ {\chi'}_{{\bf 128}, \overline{{\bf 128}}, {\bf 120}, {\bf 16}}^{{\rm SO(16)}} $ denote the characters of SO(16) which are defined by replacing  6d Sp(1) gauge theory fugacities $\{ \tilde{y}_i \}_{i\in \mathcal{I}}$ of \eqref{SO(16)ch} with the 5d Sp(2) gauge theory fugacities and instanton factor $\{ y'_{i}\}_{i\in \mathcal{I}_{{\rm 9F}}}, \mathfrak{q}'_9$ where $\mathcal{I}_{{\rm 9F}} = \{1,2,4,5,6,7,9\}$. When we set $M=0$, $\mathcal{F}'_{(1)}$ is given by $E_8$ character, so that the global symmetry is $E_8$ symmetry which is the enhanced symmetry of SU(2) gauge theory with 7 flavors discussed in \cite{Mitev:2014jza}. Therefore, we conclude that the enhanced symmetry $E_8$ is broken to SO(16) by the defects.

\begin{figure}[htb]
\centering
\includegraphics[width=11cm]{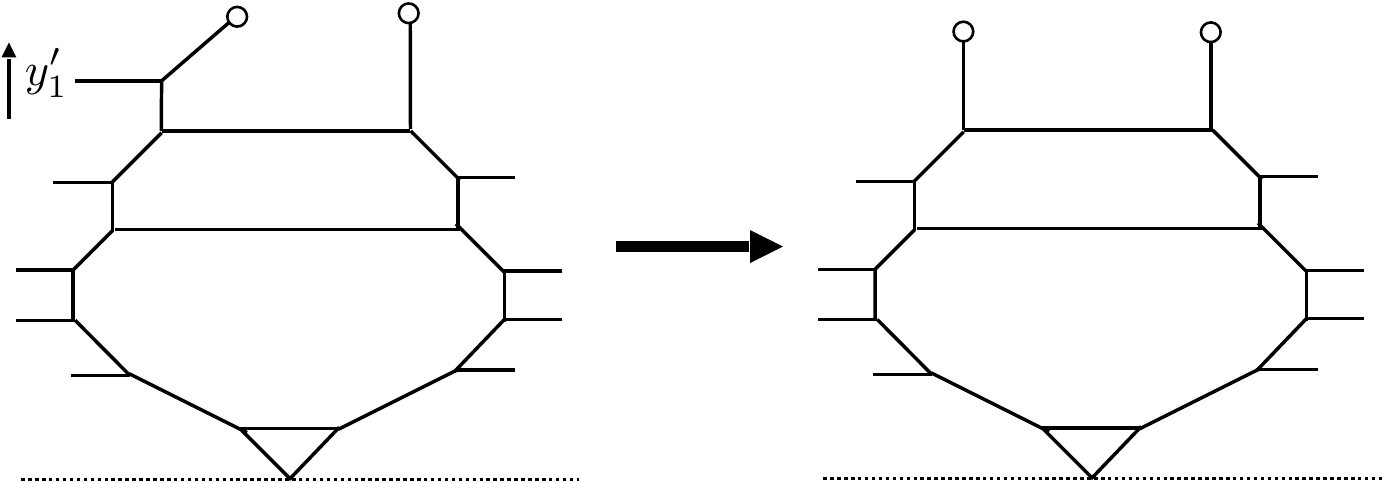}
\caption{A decoupling limit from 9 flavors to 8 flavors.}
\label{webO5_decouple2}
\end{figure}

We further consider one more flavor decoupling as in Figure \ref{webO5_decouple2} whose limit is given by
\begin{align}
&y'_{1}\to 0, \qquad
\mathfrak{q}'_9 \to 0, \qquad
\frac{{\mathfrak{q}}_9'^2}{y'_{1}}={\mathfrak{q}_{8}'}^2~ \text{(fixed)}.
\end{align}
In the same way as above case, we denote $\mathfrak{q}'_8$ as the instanton factor of 5d Sp(2) gauge theory with 8 flavors. The resulting partition function of 5d Sp(2) with 8 flavors is,
\begin{align}
{Z'}^{\rm{Sp(2)+8F}} = Z'_{(0),\text{8F}} + Z'_{(1),\text{8F}} A'_1 + \mathcal{O}({A'_1}^2),
\end{align}
with
\begin{subequations}
\begin{align}
&Z'_{(0),\text{8F}} = \frac{1}{2}  \frac{q}{(1-q)^2} \left( 2\left( {A'}_2^2 + {A'}_2^{-2} \right) -\left( {A'}_2 + {A'}_2^{-1} \right)\sum_{i=2}^{9} \left( y'_i + {y'}_i^{-1} \right)\right),
\\
&Z'_{(1),\text{8F}}=
\frac{1}{(1-q)(1-q^{-1})} 
 \frac{{\mathfrak{q}'}_8}{(1-{A'}_2^2)(1-{A'}_2^{-2})} \left(  Z'^A_{(1),\text{8F}}  + Z'^B_{(1),\text{8F}} \right),
 \end{align}
\end{subequations}
and
\begin{subequations}
\begin{align}
Z'^{{\rm SO \, SU}}_{(1),\text{8F}}
&=
-{\chi'}_{\overline{{\bf 128}}}^{{\rm SO(16)}} \chi_{1/2}^{{\rm SU(2)}}(A'_2) +2{\chi'}_{{\bf 128}}^{{\rm SO(16)}} 
\nonumber \\
&\qquad
-{\chi'}_{{\bf 16}}^{{\rm SO(16)}}\chi_{1/2}^{{\rm SU(2)}}({\mathfrak{q}'}_8)\chi_{1}^{{\rm SU(2)}}(A'_2)
+3{\chi'}_{{\bf 16}}^{{\rm SO(16)}}\chi_{1/2}^{{\rm SU(2)}}({\mathfrak{q}'}_8) ,
\\
Z'^{{\rm SU}}_{(1),\text{8F}}
&=
2\chi_{1/2}^{{\rm SU(2)}}({\mathfrak{q}'}_8) \chi_{3/2}^{{\rm SU(2)}}(A'_2)
-4\chi_{1/2}^{{\rm SU(2)}}({\mathfrak{q}'}_8) \chi_{1/2}^{{\rm SU(2)}}(A'_2).
\end{align}
\end{subequations}
Then, by defining the invariant Coulomb moduli $A'_{1,\text{inv}}$ as
\begin{align}
A'_{1,\text{inv}} ={\mathfrak{q}'}_8  A'_1,
\end{align}
the partition function has  ${\rm SO(16)}\times {\rm SU(2)}$ invariance. 

By the defect Higgsing \eqref{def5dSp2}, we have
\begin{align}
{Z'}^{{\rm SO \, SU}}_{(1),\text{8F}} \to &
-\left(q^{M}{z'} - q^{-M}{z'}^{-1} \right)\! \left(q^{M/2}{z'} - q^{-M/2}{z'}^{-1} \right) \chi_{\overline{{\bf 32}}}^{{\rm SO(12)}} 
  \nonumber \\
 &
-\left(q^{M}{z'} - q^{-M}{z'}^{-1} \right)\! \left(q^{M/2} - q^{-M/2} \right) \chi_{{\bf 32}}^{{\rm SO(12)}} \nonumber\\
&
  -\left( q^{M}{z'} - q^{-M}{z'}^{-1} \right)^2  \chi_{{\bf 12}}^{{\rm SO(12)}}\,\chi_{1/2}^{{\rm SU(2)}}({\mathfrak{q}'}_8),
\label{5dSp16F}
\end{align}
where we write the contribution only involving SO(12) characters defined by
\begin{subequations}
\begin{align}
&\chi_{{\bf 32}}^{{\rm SO(12)}} = \frac{1}{2} \left(\prod_{i\in \mathcal{I}'}({y'}_i^{1/2}+{y'}_i^{-1/2}) + \prod_{i\in \mathcal{I}'}({y'}_i^{1/2} - {y'}_i^{-1/2}) \right),
\\
&\chi_{\overline{{\bf 32}}}^{{\rm SO(12)}} = \frac{1}{2} \left(\prod_{i\in \mathcal{I}'} ({y'}_i^{1/2}+{y'}_i^{-1/2}) - \prod_{i\in \mathcal{I}'}({y'}_i^{1/2} - {y'}_i^{-1/2}) \right),
\\
&\chi_{{\bf 12}}^{{\rm SO(12)}} = \sum_{i\in \mathcal{I}'} \left( {y'}_i + {y'}_i^{-1} \right).\qquad \left( \mathcal{I}' = \{2,4,5,6,7,9 \} \right)
\end{align}
\end{subequations}
When we set $M=0$, \eqref{5dSp16F} can be expressed as $E_7$ characters,
\begin{align}
\eqref{5dSp16F} &= - \left( z' - {z'}^{-1} \right)^2 \left( \chi_{\overline{{\bf 32}}}^{\text{SO(12)}} +\chi_{1/2}^{\text{SU(2)}}(\mathfrak{q}'_8) \chi_{{\bf 12}}^{\text{SO(12)}} \right) 
=- \left( z' - {z'}^{-1} \right)^2 \chi^{E_7}_{{\bf 56}}
\end{align}


\section{Defects on pure SU(2) theories with different discrete theta angles}
\begin{figure}[htb]
\centering
\includegraphics[width=11cm]{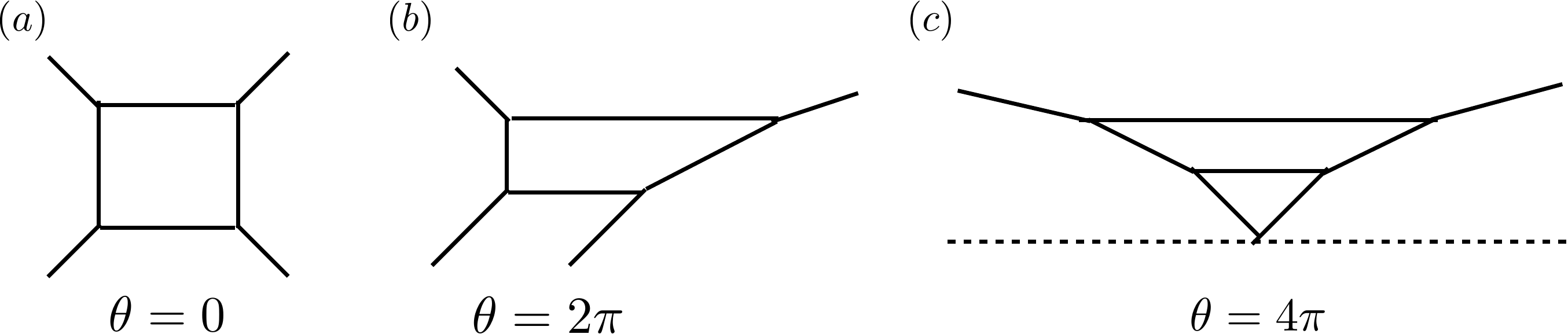}
\caption{The web diagram descriptions of pure SU(2) gauge theories with $\theta=0, 2\pi,$ and $4\pi$.}
\label{webs1}
\end{figure}

In this appendix, we explain how the defect affects the symmetry.
As an example, we shall consider 5d $\mathcal{N}=1$ pure SU(2) gauge theories with theta angle $\theta=0$, $\theta=2\pi$, and $\theta=4\pi$ whose web diagram descriptions are given in Figure \ref{webs1}.
Without the defect, these theories are equivalent through the Hanany--Witten transition.
A way to see the equivalence is to compare the partition functions of these theories.
After some computation, one can show
\begin{align}
Z_{\theta=0} = \frac{Z_{\theta=2\pi}}{Z_{E}} = Z_{\theta=4\pi},
\end{align}
where $Z_{\theta=0,2\pi,4\pi}$ denote the partition functions of pure SU(2) gauge theories with $\theta=0, 2\pi, 4\pi$, and $Z_E$ denotes the extra factor.
In this case, $Z_E$ corresponds to the strings attached between parallel external lines in Figure \ref{webs1} (b).

\begin{figure}[htb]
\centering
\includegraphics[width=8cm]{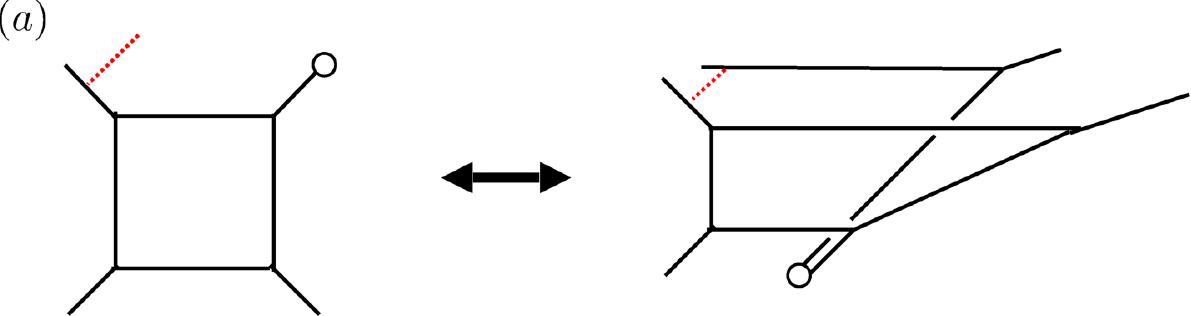}
\caption{The web diagrams after the geometric transition.}
\label{webgeom1}
\end{figure}

In the presence of the defect, however, the situation is changed: we need to include the contribution coming from the framing.
To see explicitly, let us consider following web diagrams as an example of the duality between $\theta=0$ and $2\pi$.
\begin{figure}[htb]
\centering
\includegraphics[width=10cm]{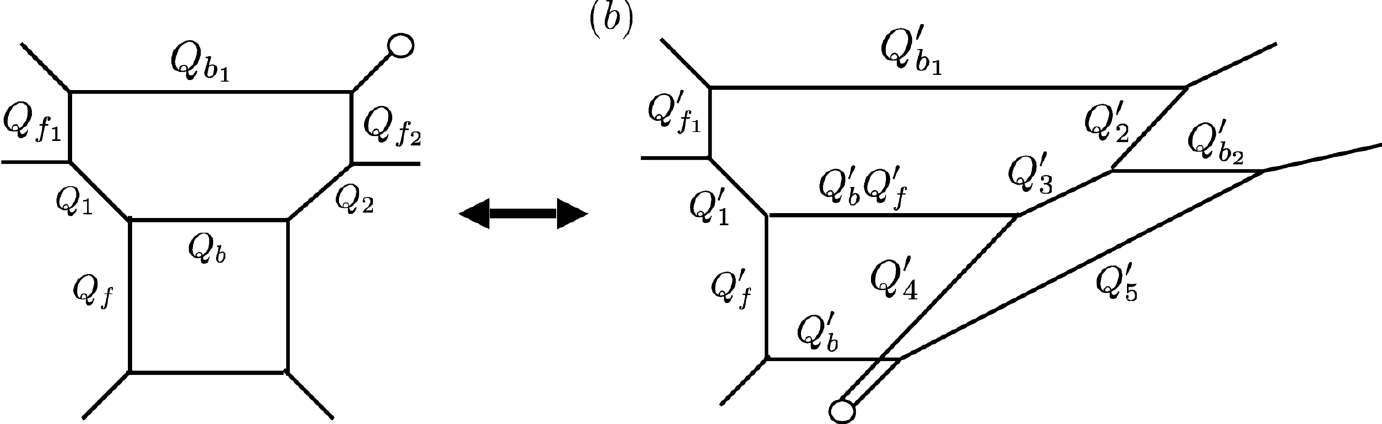}
\caption{The web diagrams which are dual each other.}
\label{webhigh1}
\end{figure}
The web diagrams (a) and (b) reduce to SU(2) gauge theory with $\theta=0$ and $\theta=2\pi$ in the presence of the defects after the geometric transition.
For simplicity, here we consider the insertion of single brane which is realized by setting the K\"ahler moduli as follows,
\begin{align}
Q_{f_1} = q^{-1},~Q_{f_2}=1,~Q'_{f_1}=q^{-1},~Q'_{3}=1.
\end{align}
Then, we find
\begin{subequations}
{\footnotesize
\begin{align}
Z^{(a)} = &
\sum_{\mu_{b_{1,2}}}(-Q_{b})^{|\mu_{b_1}|+|\mu_{b_2}|}  f_{\mu_{b_1}} f_{\mu_{b_2}}^{-1} 
s_{\mu_{b_1}}(q^{-\rho}) s_{\mu^t_{b_1}}(q^{-\rho}) s_{\mu_{b_2}}(q^{-\rho}) s_{\mu_{b_2}^t}(q^{-\rho})
\nonumber \\
&\times 
\prod_{i,j=1}^\infty \frac{1}{(1-Q_f q^{i+j-\mu_{b_1,i}-\mu_{b_2,j}^t -1})^2} 
\times
\prod_{i,j=1}^\infty \frac{(1-Q_1 q^{i+j-\mu_{b_1,j}^t-1})(1-Q_1 Q_f q^{i+j-\mu_{b_2,j}^t-1})}{(1-Q_1 q^{i+j-\mu_{b_1,j}^t-2})(1-Q_1 Q_f q^{i+j-\mu_{b_2,j}^t-2})},
\label{PFa1}
\\
Z^{(b)} = &\sum_{\mu_{b_{1,2}},\mu'_{b_{1}}} 
(-Q'_{b} {Q'_f}^2)^{|\mu_{b_1}|} (-Q'_{b})^{|\mu_{b_2}|}  f_{\mu_{b_1}}^3 f_{\mu_{b_2}} 
s_{\mu_{b_1}}(q^{-\rho}) s_{\mu^t_{b_1}}(q^{-\rho}) s_{\mu_{b_2}}(q^{-\rho}) s_{\mu_{b_2}^t}(q^{-\rho})
\nonumber \\
&\times 
\prod_{i,j=1}^\infty \frac{1}{(1-Q'_f q^{i+j-\mu_{b_1,i}-\mu_{b_2,j}^t -1})^2} 
\nonumber \\
&\times 
(-Q'_{b_1})^{|\mu'_{b_1}|} f_{\mu'_{b_1}}^3 s_{\mu'_{b_1}}(q^{-\rho})
\prod_{i,j=1}^\infty \frac{(1-Q_1 q^{i+j-\mu_{b_1,j}^t-1})(1-Q_1 Q_f q^{i+j-\mu_{b_2,j}^t-1})}{(1-Q_1 q^{i+j-\mu'_{b_1,i}-\mu_{b_1,j}^t-2})(1-Q_1 Q_f q^{i+j-\mu'_{b_1,i}-\mu_{b_2,j}^t-2})},
\label{PFb1}
\end{align}
}
\end{subequations}
where we omit the contributions which is independent of the K\"ahler moduli, and the summation of $\mu'_{b_1}$ in \eqref{PFb1} takes only the Young diagrams that have single row, $\mu'_{b_1}=\{ n \},~n\in\mathbb{Z}_{\geq 0}$.\footnote{In the following, the summation of the Young diagram with prime takes only single row.} 
Then, we check the agreement of the partition functions,
\begin{align}
Z^{(a)}  = \frac{Z^{(b)}}{Z_{\text{E},(b)}},~Z_{\text{E},(b)}=\prod_{i,j=1}^\infty \frac{1}{(1-Q'_b q^{i+j-1})},
\end{align}
under the following relations,
\begin{align}
Q_b Q_f = Q'_b,~Q_f=Q'_f,~Q_1=Q'_1,~ Q'_{b_1}=q^{-1} Q'_b Q'_f {Q'}^2_1,
\label{IDPF1}
\end{align}
as expected from Hanany--Witten transition.
Diagrammatically, the relation is expressed as in Figure \ref{webgeom1}.
The factor $(-Q'_{b_1})^{|\mu'_{b_1}|} f_{\mu'_{b_1}}^3 s_{\mu'_{b_1}}(q^{-\rho})$ comes from the trivalent diagram with the defect.

The interpretation is as follows.
The brane configuration of D5-NS5-D7-brane system are given in Table \ref{conf}.
\begin{table}[htb]
\centering
    \begin{tabular}{|c||c|c|c|c|c|c|c|c|c|c|} \hline
     IIB & $X_{0}$ & $X_{1}$ & $X_{2}$ & $X_{3}$ & $X_{4}$ & $X_{5}$ & $X_{6}$ & $X_{7}$ 
     & $X_{8}$ & $X_{9}$   \\ \hline \hline
     D5  & 
	$\circ$ & $\circ$ & $\circ$ & $\circ$ & $\circ$ & -  & - &  &  &  \\ \hline
    NS5 & 
	$\circ$ & $\circ$ & $\circ$ & $\circ$ & $\circ$ & -  & -  &  &  &  \\ \hline
    D7   &
        $\circ$ &  $\circ$ & $\circ$ & $\circ$ & $\circ$ &  &  & $\circ$ & $\circ$ & $\circ$\\ \hline
    \end{tabular}
    \caption{The brane configuration of $(p,q)$-5-brane with D7-branes. The $(p,q)$-5-brane web is defined on the $X_{5,6}$-plane denoted by bar.}
       \label{conf}
\end{table}
Since the D7-branes have branch cut and affect the charges of D5- and NS5-branes, the shape of web diagrams change by moving the D7-branes.
However, since the D7-branes do not affect the physical system, the partition function is invariant up to the extra factor. 

\begin{figure}[htb]
\centering
\includegraphics[width=10cm]{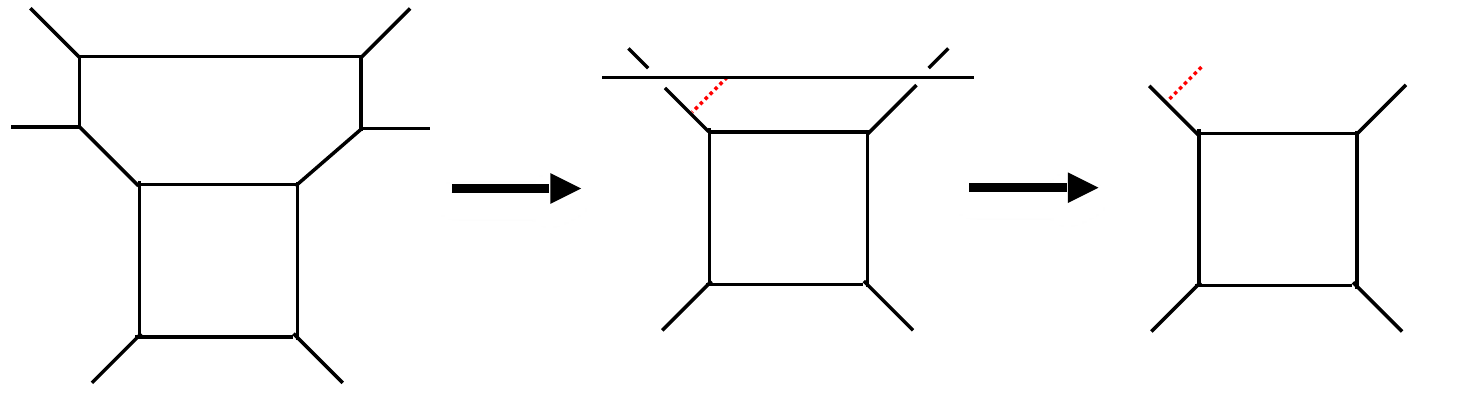}
\caption{The process of geometric transition. We can remove the framing after setting some of K\"ahler moduli.}
\label{geom2}
\end{figure}

In the topological strings, the defect corresponds to the topological brane wrapping on $S^3$ denoted by the red dashed line.
Since the partition function of A-model topological string does not depend on the complex structure moduli, we can remove the framing by taking the size of $S^3$ to infinity as one can see in Figure \ref{geom2}.

\begin{figure}[htb]
\centering
\includegraphics[width=8cm]{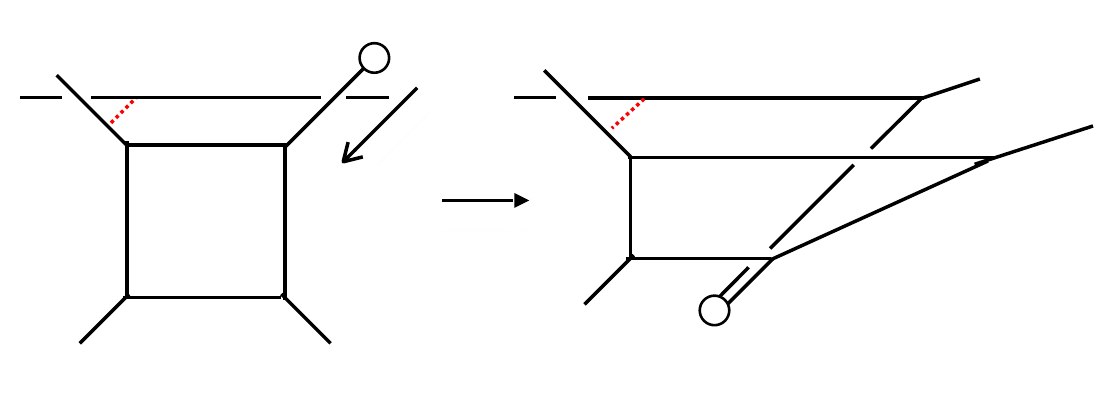}
\caption{The Hanany--Witten transition in the presence of the defect. Since the D7-brane is extended along a direction, the D7-brane affects the framing.}
\label{geom3}
\end{figure}

However, when we put the D7-branes on the web diagram, the situation changes:
Since the D7-brane is extended infinitely except for $X_{5,6}$-plane, no matter how we separate off the framing from the remaining web diagram, the framing is affected by the movement of the D7-brane, and the framing acquires the non-trivial structure which provides the non-trivial contributions.
This is why we have to consider the additional contribution coming from the framing to see the duality (see Figure \ref{geom3}).

\begin{figure}[htb]
\centering
\includegraphics[width=12cm]{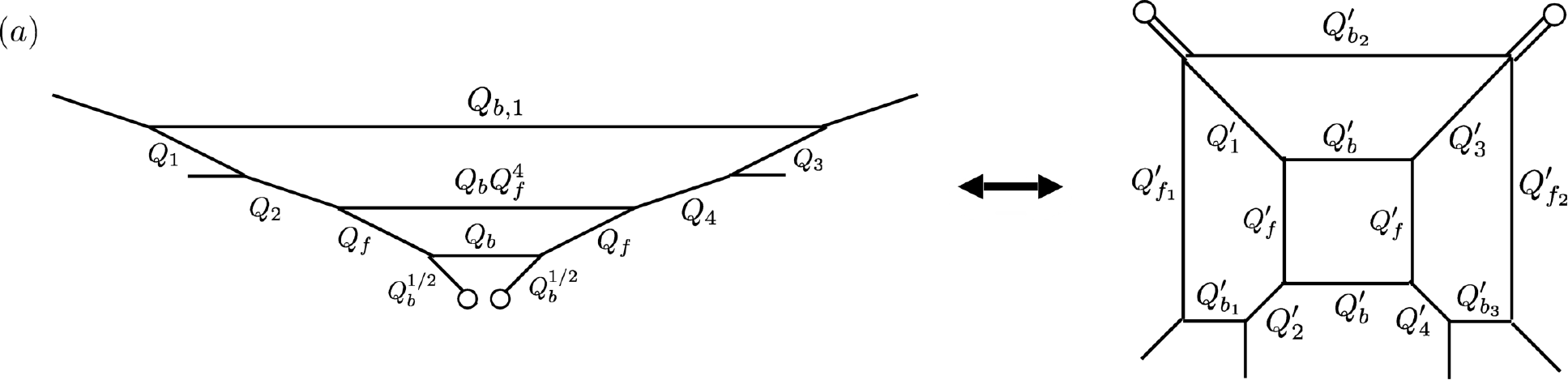}
\caption{The web diagrams which reduce to SU(2) gauge theory with $\theta=4\pi$ and $\theta=0$ in the presence of the defects after the geometric transition.}
\label{webhigh3}
\end{figure}

As another example, let us consider the duality between $\theta=0$ and $\theta=2\pi$.
Again, to find the correct duality, we utilize the geometric transition for the web diagram given in Figure \ref{webhigh3}.
Then, by setting some of the K\"ahler moduli as
\begin{align}
Q_1 =q^{-1},~Q_3=1,~Q'_{b_1}=q^{-1},~Q'_{b_3}=1,
\end{align}
we find the partition functions of the theories in the presence of single defect,
\begin{subequations}
{\footnotesize
\begin{align}
\small
Z^{(a)}
&= \sum_{\mu_{b_{1,2}}} 
(-Q_{b} Q_f^4)^{|\mu_{b_1}|} (-Q_{b})^{|\mu_{b_2}|} 
f_{\mu_{b_1}}^{-6} f_{\mu_{b_2}}^4 q^{\kappa_{\mu_{b_1}^t+\mu_{b_2}}} s^2_{\mu_{b_1}}(q^{-\rho})  s^2_{\mu^t_{b_2}}(q^{-\rho}) 
\nonumber \\
&\times 
\prod_{i,j=1}^\infty \frac{(1-Q_{b}q^{i+j-\mu_{b_2,i}-\mu_{b_2,j}-1})(1-Q_{b} Q_f q^{i+j-\mu_{b_2,i}-\mu_{b_1,j}^t-1})^2 (1-Q_{b} Q_f^2 q^{i+j-\mu_{b_1,i}^t-\mu_{b_1,j}^t-1})}{(1-Q_f q^{i+j-\mu_{b_1,i}-\mu_{b_2,j}^t -1})^2} 
\nonumber \\
&\times
\prod_{i,j=1}^\infty \frac{(1-Q_2 q^{i+j-\mu_{b_1,i}-1})(1-Q_2 Q_f q^{i+j-\mu_{b_2,j}^t-1})(1-Q_2 Q_f Q_b q^{i+j-\mu_{b_2,i}-1})(1-Q_2 Q_f^2 Q_b q^{i+j-\mu_{b_1,j}^t-1})}{(1-Q_2 q^{i+j-\mu_{b_1,i}-2})(1-Q_2 Q_f q^{i+j-\mu_{b_2,j}^t-2})(1-Q_2 Q_f Q_b q^{i+j-\mu_{b_2,i}-2})(1-Q_2 Q_f^2 Q_b q^{i+j-\mu_{b_1,j}^t-2})},
\\
Z^{(b)} &= \sum_{\mu_{f_{1,2}},\mu'_{f_{1}}} 
(-Q'_{f})^{|\mu_{f_1}|+|\mu_{f_2}|}   f_{\mu_{f_1}} f_{\mu_{f_2}}^{-1}
s_{\mu_{f_1}}(q^{-\rho}) s_{\mu^t_{f_1}}(q^{-\rho}) s_{\mu_{f_2}}(q^{-\rho}) s_{\mu_{f_2}^t}(q^{-\rho})
\nonumber \\
&\quad\times 
\prod_{i,j=1}^\infty \frac{1}{(1-Q'_b q^{i+j-\mu_{f_1,i}-\mu_{f_2,j}^t -1})^2} 
\nonumber \\
&\quad\times 
(-Q'_{f_1})^{|\mu'_{f_1}|} f_{\mu'_{f_1}}^2 s_{\mu'_{f_1}}(q^{-\rho})
\nonumber \\
&\qquad\times 
\prod_{i,j=1}^\infty \frac{(1-Q'_2 q^{i+j-\mu_{f_1,j}^t-1})(1-Q'_2 Q'_b q^{i+j-\mu_{f_2,j}^t-1})}{(1-Q'_2 q^{i+j-\mu'_{f_1,i}-\mu_{f_1,j}^t-2})(1-Q'_2 Q'_b q^{i+j-\mu'_{f_1,i}-\mu_{f_2,j}^t-2})(1-Q'_{b_2}q^{i+j-\mu'_{f_1,i}-1})}.
\end{align}
}
\end{subequations}
Then we find
\begin{align}
Z^{(a)} = \frac{Z^{(b)}}{Z_{\text{E},(b)}},~Z_{\text{E},(b)}=\prod_{i,j=1}^\infty \frac{1}{(1-{Q'}_2^2 Q'_b q^{i+j-2})},
\end{align}
under following relation,
\begin{align}
\frac{Q_b}{Q^2_f}=Q'_b,~Q_f = Q'_f,~Q_2 = Q'_2,~Q'_{f_1} =q^{-1} {Q'}^2_2 Q'_f,~Q'_{b_2} = q^{-2} {Q'}^2_2 Q'_f.
\end{align}
The corresponding web diagrams are as follows.
\begin{figure}[htb]
\centering
\includegraphics[width=8cm]{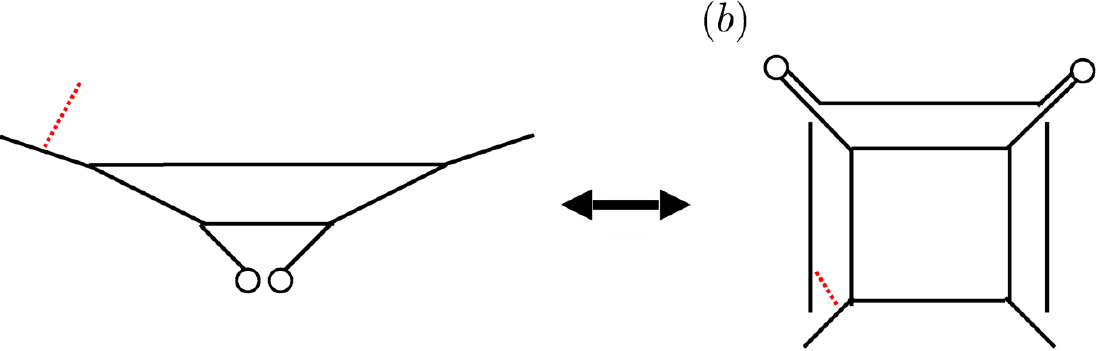}
\caption{The web diagrams after the geometric transition. Since we move two D7-branes, the shape of framing after moving the D7-branes is more complicated than previous case depicted in Figure \ref{geom3}.}
\label{webgeom3}
\end{figure}


\bigskip
\pagebreak
 
\bibliographystyle{JHEP}
\bibliography{ref}

\providecommand{\href}[2]{#2}\begingroup\raggedright\begin{thebibliography}{10}

\bibitem{Gukov:2006jk}
S.~Gukov and E.~Witten, {\it {Gauge Theory, Ramification, And The Geometric
  Langlands Program}},  \href{http://arxiv.org/abs/hep-th/0612073}{{\tt
  hep-th/0612073}}.

\bibitem{Gukov:2008sn}
S.~Gukov and E.~Witten, {\it {Rigid Surface Operators}},  {\em Adv. Theor.
  Math. Phys.} {\bf 14} (2010), no.~1 87--178,
  [\href{http://arxiv.org/abs/0804.1561}{{\tt arXiv:0804.1561}}].

\bibitem{Gomis:2007fi}
J.~Gomis and S.~Matsuura, {\it {Bubbling surface operators and S-duality}},
  {\em JHEP} {\bf 06} (2007) 025, [\href{http://arxiv.org/abs/0704.1657}{{\tt
  arXiv:0704.1657}}].

\bibitem{Alday:2009fs}
L.~F. Alday, D.~Gaiotto, S.~Gukov, Y.~Tachikawa, and H.~Verlinde, {\it {Loop
  and surface operators in N=2 gauge theory and Liouville modular geometry}},
  {\em JHEP} {\bf 01} (2010) 113, [\href{http://arxiv.org/abs/0909.0945}{{\tt
  arXiv:0909.0945}}].

\bibitem{Alday:2010vg}
L.~F. Alday and Y.~Tachikawa, {\it {Affine SL(2) conformal blocks from 4d gauge
  theories}},  {\em Lett. Math. Phys.} {\bf 94} (2010) 87--114,
  [\href{http://arxiv.org/abs/1005.4469}{{\tt arXiv:1005.4469}}].

\bibitem{Gaiotto:2012xa}
D.~Gaiotto, L.~Rastelli, and S.~S. Razamat, {\it {Bootstrapping the
  superconformal index with surface defects}},  {\em JHEP} {\bf 01} (2013) 022,
  [\href{http://arxiv.org/abs/1207.3577}{{\tt arXiv:1207.3577}}].

\bibitem{Gaiotto:2013sma}
D.~Gaiotto, S.~Gukov, and N.~Seiberg, {\it {Surface Defects and Resolvents}},
  {\em JHEP} {\bf 09} (2013) 070, [\href{http://arxiv.org/abs/1307.2578}{{\tt
  arXiv:1307.2578}}].

\bibitem{Gaiotto:2014ina}
D.~Gaiotto and H.-C. Kim, {\it {Surface defects and instanton partition
  functions}},  {\em JHEP} {\bf 10} (2016) 012,
  [\href{http://arxiv.org/abs/1412.2781}{{\tt arXiv:1412.2781}}].

\bibitem{Nazzal:2018brc}
B.~Nazzal and S.~S. Razamat, {\it {Surface Defects in E-String
  Compactifications and the van Diejen Model}},  {\em SIGMA} {\bf 14} (2018)
  036, [\href{http://arxiv.org/abs/1801.00960}{{\tt arXiv:1801.00960}}].

\bibitem{Katz:1996fh}
S.~H. Katz, A.~Klemm, and C.~Vafa, {\it {Geometric engineering of quantum field
  theories}},  {\em Nucl.Phys.} {\bf B497} (1997) 173--195,
  [\href{http://arxiv.org/abs/hep-th/9609239}{{\tt hep-th/9609239}}].

\bibitem{Katz:1997eq}
S.~Katz, P.~Mayr, and C.~Vafa, {\it {Mirror symmetry and exact solution of 4-D
  N=2 gauge theories: 1.}},  {\em Adv. Theor. Math. Phys.} {\bf 1} (1998)
  53--114, [\href{http://arxiv.org/abs/hep-th/9706110}{{\tt hep-th/9706110}}].

\bibitem{Dijkgraaf:2002fc}
R.~Dijkgraaf and C.~Vafa, {\it {Matrix models, topological strings, and
  supersymmetric gauge theories}},  {\em Nucl. Phys. B} {\bf 644} (2002) 3--20,
  [\href{http://arxiv.org/abs/hep-th/0206255}{{\tt hep-th/0206255}}].

\bibitem{Hollowood:2003cv}
T.~J. Hollowood, A.~Iqbal, and C.~Vafa, {\it {Matrix models, geometric
  engineering and elliptic genera}},  {\em JHEP} {\bf 03} (2008) 069,
  [\href{http://arxiv.org/abs/hep-th/0310272}{{\tt hep-th/0310272}}].

\bibitem{Leung:1997tw}
N.~C. Leung and C.~Vafa, {\it {Branes and toric geometry}},  {\em
  Adv.Theor.Math.Phys.} {\bf 2} (1998) 91--118,
  [\href{http://arxiv.org/abs/hep-th/9711013}{{\tt hep-th/9711013}}].

\bibitem{Dimofte:2010tz}
T.~Dimofte, S.~Gukov, and L.~Hollands, {\it {Vortex Counting and Lagrangian
  3-manifolds}},  {\em Lett.Math.Phys.} {\bf 98} (2011) 225--287,
  [\href{http://arxiv.org/abs/1006.0977}{{\tt arXiv:1006.0977}}].

\bibitem{Awata:2010bz}
H.~Awata, H.~Fuji, H.~Kanno, M.~Manabe, and Y.~Yamada, {\it {Localization with
  a Surface Operator, Irregular Conformal Blocks and Open Topological String}},
   {\em Adv. Theor. Math. Phys.} {\bf 16} (2012), no.~3 725--804,
  [\href{http://arxiv.org/abs/1008.0574}{{\tt arXiv:1008.0574}}].

\bibitem{Gopakumar:1998ki}
R.~Gopakumar and C.~Vafa, {\it {On the gauge theory / geometry
  correspondence}},  {\em Adv. Theor. Math. Phys.} {\bf 3} (1999) 1415--1443,
  [\href{http://arxiv.org/abs/hep-th/9811131}{{\tt hep-th/9811131}}].

\bibitem{Ooguri:1999bv}
H.~Ooguri and C.~Vafa, {\it {Knot invariants and topological strings}},  {\em
  Nucl. Phys. B} {\bf 577} (2000) 419--438,
  [\href{http://arxiv.org/abs/hep-th/9912123}{{\tt hep-th/9912123}}].

\bibitem{Mori:2016qof}
H.~Mori and Y.~Sugimoto, {\it {Surface Operators from M-strings}},  {\em Phys.
  Rev.} {\bf D95} (2017), no.~2 026001,
  [\href{http://arxiv.org/abs/1608.02849}{{\tt arXiv:1608.02849}}].

\bibitem{Haghighat:2013gba}
B.~Haghighat, A.~Iqbal, C.~Koz{\c c}az, G.~Lockhart, and C.~Vafa, {\it
  {M-Strings}},  {\em Commun. Math. Phys.} {\bf 334} (2015), no.~2 779--842,
  [\href{http://arxiv.org/abs/1305.6322}{{\tt arXiv:1305.6322}}].

\bibitem{Sugimoto:2015nha}
Y.~Sugimoto, {\it {The Enhancement of Supersymmetry in M-strings}},  {\em Int.
  J. Mod. Phys.} {\bf A31} (2016), no.~16 1650088,
  [\href{http://arxiv.org/abs/1508.02125}{{\tt arXiv:1508.02125}}].

\bibitem{Kim:2015jba}
S.-S. Kim, M.~Taki, and F.~Yagi, {\it {Tao Probing the End of the World}},
  {\em PTEP} {\bf 2015} (2015), no.~8 083B02,
  [\href{http://arxiv.org/abs/1504.03672}{{\tt arXiv:1504.03672}}].

\bibitem{Aganagic:2003db}
M.~Aganagic, A.~Klemm, M.~Marino, and C.~Vafa, {\it {The Topological vertex}},
  {\em Commun.Math.Phys.} {\bf 254} (2005) 425--478,
  [\href{http://arxiv.org/abs/hep-th/0305132}{{\tt hep-th/0305132}}].

\bibitem{Kim:2017jqn}
S.-S. Kim and F.~Yagi, {\it {Topological vertex formalism with O5-plane}},
  {\em Phys. Rev.} {\bf D97} (2018) 026011,
  [\href{http://arxiv.org/abs/1709.01928}{{\tt arXiv:1709.01928}}].

\bibitem{Gaiotto:2015una}
D.~Gaiotto and H.-C. Kim, {\it {Duality walls and defects in 5d $ \mathcal{N}=1
  $ theories}},  {\em JHEP} {\bf 01} (2017) 019,
  [\href{http://arxiv.org/abs/1506.03871}{{\tt arXiv:1506.03871}}].

\bibitem{Hayashi:2015fsa}
H.~Hayashi, S.-S. Kim, K.~Lee, M.~Taki, and F.~Yagi, {\it {A new 5d description
  of 6d D-type minimal conformal matter}},  {\em JHEP} {\bf 08} (2015) 097,
  [\href{http://arxiv.org/abs/1505.04439}{{\tt arXiv:1505.04439}}].

\bibitem{Yun:2016yzw}
Y.~Yun, {\it {Testing 5d-6d dualities with fractional D-branes}},  {\em JHEP}
  {\bf 12} (2016) 016, [\href{http://arxiv.org/abs/1607.07615}{{\tt
  arXiv:1607.07615}}].

\bibitem{Kim:2014dza}
J.~Kim, S.~Kim, K.~Lee, J.~Park, and C.~Vafa, {\it {Elliptic Genus of
  E-strings}},  \href{http://arxiv.org/abs/1411.2324}{{\tt arXiv:1411.2324}}.

\bibitem{Hayashi:2016abm}
H.~Hayashi, S.-S. Kim, K.~Lee, and F.~Yagi, {\it {Equivalence of several
  descriptions for 6d SCFT}},  {\em JHEP} {\bf 01} (2017) 093,
  [\href{http://arxiv.org/abs/1607.07786}{{\tt arXiv:1607.07786}}].

\bibitem{Brunner:1997gf}
I.~Brunner and A.~Karch, {\it {Branes at orbifolds versus Hanany Witten in
  six-dimensions}},  {\em JHEP} {\bf 9803} (1998) 003,
  [\href{http://arxiv.org/abs/hep-th/9712143}{{\tt hep-th/9712143}}].

\bibitem{Hanany:1997gh}
A.~Hanany and A.~Zaffaroni, {\it {Branes and six-dimensional supersymmetric
  theories}},  {\em Nucl.Phys.} {\bf B529} (1998) 180--206,
  [\href{http://arxiv.org/abs/hep-th/9712145}{{\tt hep-th/9712145}}].

\bibitem{Sen:1996vd}
A.~Sen, {\it {F theory and orientifolds}},  {\em Nucl.Phys.} {\bf B475} (1996)
  562--578, [\href{http://arxiv.org/abs/hep-th/9605150}{{\tt hep-th/9605150}}].

\bibitem{Hwang:2014uwa}
C.~Hwang, J.~Kim, S.~Kim, and J.~Park, {\it {General instanton counting and 5d
  SCFT}},  {\em JHEP} {\bf 07} (2015) 063,
  [\href{http://arxiv.org/abs/1406.6793}{{\tt arXiv:1406.6793}}]. [Addendum:
  JHEP04,094(2016)].

\bibitem{Hayashi:2017btw}
H.~Hayashi, S.-S. Kim, K.~Lee, and F.~Yagi, {\it {Discrete theta angle from an
  O5-plane}},  {\em JHEP} {\bf 11} (2017) 041,
  [\href{http://arxiv.org/abs/1707.07181}{{\tt arXiv:1707.07181}}].

\bibitem{Hayashi:2015vhy}
H.~Hayashi, S.-S. Kim, K.~Lee, M.~Taki, and F.~Yagi, {\it {More on 5d
  descriptions of 6d SCFTs}},  {\em JHEP} {\bf 10} (2016) 126,
  [\href{http://arxiv.org/abs/1512.08239}{{\tt arXiv:1512.08239}}].

\bibitem{Hayashi:2015zka}
H.~Hayashi, S.-S. Kim, K.~Lee, and F.~Yagi, {\it {6d SCFTs, 5d Dualities and
  Tao Web Diagrams}},  \href{http://arxiv.org/abs/1509.03300}{{\tt
  arXiv:1509.03300}}.

\bibitem{Cheng:2018wll}
S.~Cheng and S.-S. Kim, {\it {Refined topological vertex for 5d $Sp(N)$ gauge
  theories with antisymmetric matter}},
  \href{http://arxiv.org/abs/1809.00629}{{\tt arXiv:1809.00629}}.

\bibitem{Kim:2015fxa}
J.~Kim, S.~Kim, and K.~Lee, {\it {Higgsing towards E-strings}},
  \href{http://arxiv.org/abs/1510.03128}{{\tt arXiv:1510.03128}}.

\bibitem{Taki:2010bj}
M.~Taki, {\it {Surface Operator, Bubbling Calabi-Yau and AGT Relation}},  {\em
  JHEP} {\bf 1107} (2011) 047, [\href{http://arxiv.org/abs/1007.2524}{{\tt
  arXiv:1007.2524}}].

\bibitem{Ruijsenaars_2015}
S.~N. Ruijsenaars, {\it Hilbert-schmidt operators vs. integrable systems of
  elliptic calogero-moser type iv. the relativistic heun (van diejen) case},
  {\em Symmetry, Integrability and Geometry: Methods and Applications} (Jan,
  2015) [\href{http://arxiv.org/abs/1404.4392}{{\tt arXiv:1404.4392}}].

\bibitem{Haghighat:2018dwe}
B.~Haghighat, J.~Kim, W.~Yan, and S.-T. Yau, {\it {D-type fiber-base duality}},
   {\em JHEP} {\bf 09} (2018) 060, [\href{http://arxiv.org/abs/1806.10335}{{\tt
  arXiv:1806.10335}}].

\bibitem{Chen:2020jla}
J.~Chen, B.~Haghighat, H.-C. Kim, and M.~Sperling, {\it {Elliptic Quantum
  Curves of Class $\mathcal{S}_k$}},
  \href{http://arxiv.org/abs/2008.05155}{{\tt arXiv:2008.05155}}.

\bibitem{Kimura:2018kaf}
T.~Kimura and Y.~Sugimoto, {\it {Quantum mirror curve of periodic chain
  geometry}},  {\em JHEP} {\bf 04} (2019) 147,
  [\href{http://arxiv.org/abs/1810.01885}{{\tt arXiv:1810.01885}}].

\bibitem{Iqbal:2004ne}
A.~Iqbal and A.-K. Kashani-Poor, {\it {The Vertex on a strip}},  {\em
  Adv.Theor.Math.Phys.} {\bf 10} (2006) 317--343,
  [\href{http://arxiv.org/abs/hep-th/0410174}{{\tt hep-th/0410174}}].

\bibitem{Konishi:2006ev}
Y.~Konishi and S.~Minabe, {\it {Flop invariance of the topological vertex}},
  {\em Int.J.Math.} {\bf 19} (2008) 27--45,
  [\href{http://arxiv.org/abs/math/0601352}{{\tt math/0601352}}].

\bibitem{Taki:2008hb}
M.~Taki, {\it {Flop Invariance of Refined Topological Vertex and Link
  Homologies}},  \href{http://arxiv.org/abs/0805.0336}{{\tt arXiv:0805.0336}}.

\bibitem{Mitev:2014jza}
V.~Mitev, E.~Pomoni, M.~Taki, and F.~Yagi, {\it {Fiber-Base Duality and Global
  Symmetry Enhancement}},  {\em JHEP} {\bf 1504} (2015) 052,
  [\href{http://arxiv.org/abs/1411.2450}{{\tt arXiv:1411.2450}}].

\end{thebibliography}\endgroup
\end{document}